\documentclass[12pt,a4paper]{article}
\usepackage{a4}
\newcommand{\abc}[1]{\mbox{#1)}\quad}
\newcommand{\bm}[1]{\mbox{\boldmath $#1$}}
\newcommand{\DD}{\mbox{\tiny $||$}}
\newcommand{\deriv}[2]{\mbox{$\displaystyle \frac{{\rm d}#1}{{\rm d}#2}$}}
\newcommand{\ds}{\displaystyle}

\newcommand{\Li}[2]{\mbox{$\pounds_{\mbox{\footnotesize $#1$}}{#2}$}}
\newcommand{\pderiv}[2]{\mbox{$\displaystyle\frac{{\partial}#1}{{\partial}#2}$}}
\renewcommand{\,}{{\!}}

\def\om{\stackrel{4}{\omega}{}\!\!}
\def\ge{\stackrel{4}{g}{}\!\!}
\def\riv{\stackrel{4}{R}{}\!\!}
\def\rif{\stackrel{5}{R}{}\!\!}
\newcommand{\CR}[2]{\Biggl\{\matrix{#1\cr#2\cr}\Biggr\}}
\newcommand{\nabv}[1]{\mbox{$\stackrel{4}{\nabla}_{\mbox{\footnotesize $#1$}}$}}
\newcommand{\nabf}[1]{\mbox{$\stackrel{5}{\nabla}_{\mbox{\footnotesize $#1$}}$}}
\newcommand{\til}[1]{\stackrel{\sim}{#1}{\!}}{}

\def\newpic#1{}

\newcommand{\mathscr}[1]{{\cal #1}}
\savebox1{
\unitlength 1mm
\linethickness{0.6pt}
\begin{picture}(96.00,88.33)
\bezier{268}(20.00,25.00)(49.67,33.00)(85.00,25.00)
\bezier{104}(20.00,25.00)(28.00,37.00)(20.00,45.00)
\bezier{284}(20.00,45.00)(53.00,58.33)(86.00,46.00)
\bezier{128}(86.00,46.00)(74.00,36.33)(86.00,25.00)
\bezier{168}(33.00,37.00)(68.67,35.00)(70.00,41.00)
\bezier{268}(30.00,55.00)(59.67,63.00)(95.00,55.00)
\bezier{104}(30.00,55.00)(38.00,67.00)(30.00,75.00)
\bezier{284}(30.00,75.00)(63.00,88.33)(96.00,76.00)
\bezier{128}(96.00,76.00)(84.00,66.33)(96.00,55.00)
\bezier{168}(43.00,67.00)(78.67,65.00)(80.00,71.00)
\bezier{44}(43.00,76.00)(42.67,84.33)(44.00,87.00)
\bezier{60}(30.00,17.00)(28.00,24.00)(30.00,31.00)
\bezier{52}(30.00,31.00)(33.00,40.00)(32.00,37.00)
\bezier{80}(32.00,37.00)(37.50,46.00)(37.00,55.00)
\bezier{96}(37.00,55.00)(44.00,66.00)(43.00,77.00)
\bezier{44}(57.00,76.00)(56.67,84.33)(58.00,87.00)
\bezier{44}(71.00,76.00)(70.67,84.33)(72.00,87.00)
\bezier{60}(44.00,17.00)(42.00,24.00)(44.00,31.00)
\bezier{60}(58.00,17.00)(56.00,24.00)(58.00,31.00)
\bezier{52}(44.00,31.00)(47.00,40.00)(46.00,37.00)
\bezier{52}(58.00,31.00)(61.00,40.00)(60.00,37.00)
\bezier{80}(46.00,37.00)(51.50,46.00)(51.00,55.00)
\bezier{80}(60.00,37.00)(65.50,46.00)(65.00,55.00)
\bezier{96}(51.00,55.00)(58.00,66.00)(57.00,77.00)
\bezier{96}(65.00,55.00)(72.00,66.00)(71.00,77.00)
\bezier{60}(69.00,21.00)(67.00,28.00)(69.00,35.00)
\bezier{52}(69.00,35.00)(72.00,44.00)(71.00,41.00)
\bezier{80}(71.00,41.00)(76.50,50.00)(76.00,59.00)
\bezier{96}(76.00,59.00)(83.00,70.00)(82.00,81.00)
\bezier{252}(32.00,37.00)(67.56,50.00)(81.00,71.00)
\put(77.00,30.00){\makebox(0,0)[cc]{$\tau (X^{\alpha})=0$}}
\put(85.00,60.00){\makebox(0,0)[cc]{$\tau = \tau_1$}}
\put(81.00,71.00){\circle*{2.00}}
\put(32.00,37.00){\circle*{2.00}}
\put(37.00,55.00){\circle*{2.00}}
\put(27.00,36.00){\makebox(0,0)[cc]{$P'$}}
\put(34.00,53.00){\makebox(0,0)[cc]{$P_2$}}
\put(38.00,66.00){\makebox(0,0)[cc]{$P_1$}}
\put(42.00,67.00){\circle*{2.00}}
\put(56.00,38.00){\makebox(0,0)[cc]{$\lambda ' (t)$}}
\put(57.00,40.00){\vector(-1,4){1.75}}
\put(47.00,20.00){\makebox(0,0)[cc]{$\gamma (t)$}}
\put(48.00,22.00){\vector(1,4){3.50}}
\put(62.00,78.00){\makebox(0,0)[cc]{$\lambda (t)$}}
\put(62.00,77.00){\vector(0,-1){10.00}}
\put(30.00,11.00){\makebox(0,0)[cc]{$X^{\alpha}(x^m_0,\tau)$}}
\put(58.00,11.00){\makebox(0,0)[cc]{$X^{\alpha}(x^m,\tau)$}}
\put(71.00,41.00){\circle*{2.00}}
\end{picture}
}
\begin{document}%

\baselineskip=18pt

\title{On the Axiomatics of the 5-dimensional
Projective Unified Field Theory of Schmutzer}
\author{A.K.Gorbatsievich\thanks{
On leave of absence from the  Department of Theoretical Physics of the
          Byelorussian State University,
          Minsk 220080, Belarus (E-mail:
          gorbatsievich@phys.bsu.unibel.by).
    The author is very grateful to DAAD and FSU Jena (Germany) for
    financial support and hospitality.}~~\thanks{Dedicated to
my academic teacher Prof. Dr. Ernst Schmutzer on the occasion
of his $70^{\rm th}$ birthday.}
\\[0.5ex]
       Theoretisch-Physikalisches Institut\\
       Friedrich-Schiller-Universit\"at Jena\\
       D-07743 Jena, Germany
}
\date{}

\maketitle
\thispagestyle{empty}
\vspace{-1ex}
\centerline{(Received March 11, 2000)}

\textbf{Abstract}\\[1ex]
{\small For more than 40 years E.Schmutzer has developed a new
approach to the (5-dimensional) projective relativistic theory
which he later called Projective Unified Field Theory (PUFT). In
the present paper we introduce a new axiomatics for Schmutzer's
theory. By means of this axiomatics we can give a new geometrical
interpretation of the physical concept of the PUFT.
\\[2ex]
}

\section {Introduction}
As it is well known the 5-dimensional idea of a unified field
theory goes back to the works of Kaluza and Klein
\cite{Kaluza,Klein}. The pioneers of the projective approach to
this theory were Veblen and van Dantzig \cite{Dantzig,Veblen}.
Later this approach was developed further by many other authors.

An essential progress in this projective type of theories was done
by Jordan \cite {Jordan} who first took into consideration the
occurring scalar field which inevitably appears in this theory.
However, the field equations used by him were unacceptable.

A basically different approach to a projective field theory was
proposed by E.Schmutzer \cite {Schmutzer1} who (according to the
requirements of a unified field theory) developed further and
applied a basis vectors formalism initiated by Hessenberg, Schouten
and others \cite {Schouten} in the theory of manifolds. He had no
longer considered the scalar field mentioned above to be an
auxiliary one. On the contrary he associated this field with a new
phenomenon of nature being on one the level of gravitation and
electromagnetism. In 1980 he introduced the new term
\textbf{``scalarism''} \cite{Schmutzer2} for this phenomenon.
For this hypothetically used scalar field with fundamental
importance in the PUFT Schmutzer introduced the term
\textbf{``scalaric field''} in order to distinguish it from the
various other scalar fields in physics. The most interesting and
important results of application of PUFT are presented in the
Appendix.

Beside the projective relativistic theory many authors were
actively developing further the initial Kaluza - Klein theory
aiming at a unified field theory of elementary particles. Here we
only refer to the monographs of Wesson \cite{Wesson1} and
Vladimirov \cite{Vladimirov}, where one can find references to the
historical, mathematical and physical literature on this subject.

Concluding this introduction we would like to mention the new
5-dimensional original field theory by Wesson
\cite{Wesson,Wesson1,Wesson2} recently appeared and offered for
discussion.

In the first two versions of PUFT (see \cite {Schmutzer1} and \cite
{Schmutzer2} respectively) Schmutzer used 5-dimensional
Einstein-like field equations.\footnote{Greek indices run from 1 to
5, Latin indices from 1 to 4; the signatures are: of the
5-dimensional metric $(+++-+)$, of the space-time metric $(+++-)$.
Comma means the partial and semicolon the covariant derivative,
respectively.}
\begin{eqnarray} \label{W1}
    \stackrel{5}{R}_{\mu\nu}-
    \frac{1}{2}g_{\mu\nu}\stackrel{5}{R}{}+
         \Lambda _{\mu \nu }=\kappa_0
    \Theta_{\mu\nu}
\end{eqnarray}
Afterwards with the help of a special projection
procedure (details can be found in the papers quoted above) a
system of 4-dimensional field equations describing gravitation,
electromagnetism and scalarism was derived. Here
$\ds\kappa_{0} = \frac{8\pi G}{c^{4}}$ is Einstein's
gravitational constant,
       $\theta_{\alpha\varepsilon}$
       is the so-called energy projector of the non-geometrized
       matter named ``\textbf{substrate}'',
       $\rif_{\alpha\varepsilon }$ is the 5-dimensional Ricci
       tensor, $\rif$ is the 5-dimensional curvature invariant, and
\begin{eqnarray}\label{z2}
 \abc{a} \Lambda _{\mu \nu }=\lambda _0 g_{\mu \nu } \quad
     \mbox{resp.}\quad
 \abc{b} \Lambda _{\mu \nu }=\lambda _0 (g_{\mu \nu }+s_\mu s_\nu)S
 \end{eqnarray}
are analogs of the cosmological terms in Version I and Version
II, respectively. Here $g _ {\mu \nu} $ is the metric tensor. In a
special frame $
\left\{X ^ {\mu} \right\} $ the unit vector $s ^ {\mu} $ has
(see the next section for details) the following form $\ds s ^
{\mu} = \frac {X ^ {\mu}} {S} $, where $S =\sqrt {X _ {\mu} X ^
{\mu}} =S_0e ^\sigma $ ($S_0 $ is an arbitrary constant
of the dimension of length).

The 5-dimensional Ricci tensor and the 5-dimensional curvature
invariant, both mentioned above, are defined as follows: $
\rif_{\alpha\varepsilon } = \rif^{\tau}\,_{\alpha\varepsilon \tau},
    \quad \rif = \rif^{\alpha}\,_{\alpha},
$
where
\begin{eqnarray}\label{z1}
\rif^{\alpha}\,_{\mu\nu\varepsilon } \equiv\>
    \CR{\alpha}{\mu\varepsilon }_{,\nu} -
    \CR{\alpha}{\mu\nu }_{,\varepsilon } +
    \CR{\tau}{\mu\varepsilon }\CR{\alpha}{\tau\nu}
    - \CR{\tau}{\mu\nu }\CR{\alpha}{\tau\varepsilon }.
\end{eqnarray}

For physical reasons in the following the Gauss system of units is
chosen.

\underline{\bf  Version I.}
  The 4-dimensional field equations
  (without a cosmological term: $\Lambda _{\mu\nu} = 0$)
 have the following form:
     \begin{eqnarray} \label{th.2}
           \riv_{mn}-{1\over 2}\ge_{mn}\riv
          =\kappa(E_{mn}+S_{mn}+\Theta_{mn}).
\end{eqnarray}
These equations are generalized 4-dimensional equations for the
gravitational field, where $\ds \kappa = \kappa_{0} e^{-\sigma}$,
\begin{eqnarray} \label{th.3}
E_{mn}={1\over 4\pi}(B_{mk}H^{k}\,_{n}+{1\over 4}g_{mn}B_{jk}H^{jk})
\end{eqnarray}
is the electromagnetic energy--momentum tensor and
\begin{eqnarray} \label{th.4a}
    S_{mn} = - \frac{1}{2\kappa}
    (\sigma_{,m}\sigma_{,n}
    + \sigma_{,m;n}
    - g_{mn}(\sigma_{,k}\sigma^{,k}
    + \sigma^{,k}\,_{;k})
\end{eqnarray}
is the energy--momentum tensor of the scalaric field $\sigma$.
Further following field equations hold:
\begin{eqnarray} \label{th.5a}
           a)\; H^{mn}\,_{;n}={4\pi\over c}j^{m} , \qquad
          b)\; B_{[mn,k]}=0, \qquad
          c)\; H_{mn}=e^{3\sigma}B_{mn},
\end{eqnarray}
\begin{eqnarray} \label{th.6a}
           \sigma^{,k}\,_{;k}=
          \kappa_{0}\left(\frac{2}{3}
          \vartheta+{1\over 8\pi}B_{kj}H^{kj} \right).
\end{eqnarray}
These are the electromagnetic field equations and the field
equation for the scalaric field $\sigma$. Here the following
notations were used: $\riv_{mn}$, $\riv$ are the Ricci tensor and
the curvature invariant in the 4-dimensional space-time,
respectively. $H^{mn}$, $B^{mn}$ are the electromagnetic induction
tensor and the electromagnetic field strength tensor, respectively.
The quantity $\vartheta$ being one of the sourses of the scalaric
field is called scalaric substrate energy density.

The idea of developing the \underline{\bf Version II }
\cite{Schmutzer2} was to remove the second order derivatives in the
energy -- momentum tensor (\ref{th.4a}) of the scalaric field. By
means of a modified projection formalism it became possible to
obtain a system of equations being slightly different from the
analogous system of the version I, given by the equations
(\ref{th.4a}), (\ref{th.5a}c) and (\ref{th.6a}), namely:
\begin{eqnarray}\label{z3}
     \riv_{mn}-\frac{1}{2}\riv\ge_{mn}+\lambda  _0 S_0\ge_{mn}=
     \kappa _0(\theta _{mn} +E_{mn}+S_{mn}),
\end{eqnarray}
\begin{eqnarray} \label{th.4}
    S_{mn} = - \frac{3}{2\kappa_0}
    (\sigma_{,m}\sigma_{,n}-{1\over 2}g_{mn}
    \sigma_{,k}\sigma^{,k}),
\end{eqnarray}
\begin{eqnarray} \label{th.5}
           a)\; H^{mn}\,_{;n}={4\pi\over c}j^{m} , \qquad
          b)\; B_{[mn,k]}=0, \qquad
          c)\; H_{mn}=e^{3\sigma}B_{mn},
\end{eqnarray}
\begin{eqnarray} \label{th.6}
           \sigma^{,k}\,_{;k}=
          \kappa_{0}\left(\frac{2}{3}
          \vartheta+{1\over 8\pi}B_{kj}H^{kj} \right).
\end{eqnarray}

However, the new projection formalism led to other problems,
particularly in the spinor theory. Therefore approximately in 1994
E.Schmutzer left this version II and offered version III.

In the \textbf{version III} by deeply founded considerations on the
level of the Lagrange-Hamilton formalism the following new
5-dimensional field equations were found \cite{Schmutzer5}:
\begin{eqnarray} \label{W17}
  & & R_{\mu\nu}-{1\over 2} g_{\mu\nu} \stackrel{5}{R} -
  {1\over S} S_{,\mu ;\nu} +
  {K_0 \kappa _0 \over {S^2}} S_{,\mu}S_{,\nu}- {1\over S}
  s_{\mu}s_{\nu} [2(1- {1\over 2}K_0 \kappa_0)S^{,\tau}_{\ ;\tau}  \nonumber\\
  & & -{3 \over S} (1- {1 \over 2}
  K_0 \kappa_0) S_{,\tau} S^{,\tau} + {3 \lambda_0 \over S}- {S \over
  2} \stackrel{5}{R} ]+ {1 \over S} g_{\mu \nu}
  [S^{,\tau}_{\ ;\tau}  \nonumber\\
  & & -{1 \over S} (1+ {1 \over 2} K_0 \kappa_0) S_{,\tau} S^{,\tau}+ {1 \over S}
  \lambda_0 ]  =  \kappa_0 \Theta_{\mu\nu}
\end{eqnarray}
    ($\lambda _0$  is  a  kind of cosmological  constant.
$K_0$ is a free constant, where Schmutzer preferred the choice
$K_0=-2$). Compared with the Einstein-like field
    equation (\ref{W1}) this is a rather complicated equation, but
    it  fulfils  important physical  demands    mentioned    in    the
    \cite{Schmutzer5}. From (\ref{W17}) one  obtains
\begin{eqnarray} \label{h.2}
           R_{mn}-{1\over 2}g_{mn}R +
                          \frac{\lambda _0}{S_0^2}
                          e^{-2\sigma}g_{mn}
          =\kappa_{0}(E_{mn}+S_{mn}+\Theta_{mn}),
\end{eqnarray}
\begin{eqnarray} \label{h.3}
           E_{mn}={1\over 4\pi}(B_{mk}H^{k}\,_{n}+{1\over 4}g_{mn}B_{jk}H^{jk}),
\end{eqnarray}
\begin{eqnarray} \label{h.4}
    S_{mn} = - K_{0}
    (\sigma_{,m}\sigma_{,n}-{1\over 2}g_{mn}
    \sigma_{,k}\sigma^{,k}),
\end{eqnarray}
\begin{eqnarray} \label{h.5}
          a)\; H^{mn}\,_{;n}={4\pi\over c}j^{m} , \qquad
          b)\; B_{[mn,k]}=0, \qquad
          c)\; H_{mn}=e^{2\sigma}B_{mn},
\end{eqnarray}
\begin{eqnarray} \label{h.6}
           \sigma^{,k}\,_{;k}= \frac{1}{K_{0}}\left(
          \vartheta+{1\over 8\pi}B_{kj}H^{kj} \right)
          + \frac{2\lambda_0}{\kappa_{0} K_{0}S_0^2}e^{-2\sigma}.
\end{eqnarray}

In the present paper we introduce a new geometrical axiomatics for
the Schmutzer theory. By means of this axiomatics we can give a new
geometrical interpretation of physical results obtained in the
PUFT.

\section{Projection formalism}

As it is well known, the physical basis of the 5-dimensional
Projective Unified Field Theory is supported by the following
mathematical theorem: The semidirect product of the Abelian group
of gauge transformations (electromagnetism) and of the group of the
general 4-dimensional coordinate transformations (gravitation)
corresponds to the group being homomorphic to the group of all
5-dimensional homogeneous of degree 1 coordinate transformations
\begin{eqnarray}\label{2.1}
         X^{\mu '} = X^{\mu '}(X^{\nu}) = \frac{1}{\alpha}
         X^{\mu '}(\alpha X^{\nu}) \qquad (\alpha={\rm const}).
\end{eqnarray}
This mathematical theorem allows us to assume that the geometry,
constructed on this group, can be a basis for the geometrization of
the electromagnetic, the gravitational and the scalaric field.

From the equation (\ref{2.1}) and Euler's theorem on homogeneous
functions follows that these special coordinates $X^{\mu}$ in the
5-dimensional space $\mathscr{M}_{5}$ are transformed as the
components of a vector:
\begin{eqnarray} \label{2.2}
         X^{\mu '} = X^{\mu '}\,_{,\nu} X^{\nu}.
\end{eqnarray}
Further the vector
\begin{eqnarray} \label{2.3}
      \bm{\cal  R}=X^{\mu}\bm{E}_{\mu},\quad
           \bm{E}_{\mu}    = \pderiv{~}{X^{\mu}}
\end{eqnarray}
can be regarded as 5-dimensional radius vector. This was a very
important starting point of Schmutzer in 1957. Also in the
following this vector field \bm{\cal R} plays a fundamental role.
Of course, it is possible to introduce in the space
$\mathscr{M}_{5}$ arbitrary coordinates $y^{\mu} =
y^{\mu}(X^{\alpha})$.

In context with the theorem mentioned above we should remark that
the 4-dimensional coordinates $\{x^{i}\}$ in the space-time should
satisfy the equation
\begin{eqnarray} \label{2.4}
         x^{i}\,_{,\nu}X^{\nu} = 0
\end{eqnarray}
(for details see \cite{Schmutzer1}).

In order to construct a projection formalism, let us
consider the congruence
\begin{eqnarray} \label{th.8}
         y^{\nu} = y^{\nu}(x^i, \tau)
\end{eqnarray}
of integral curves of the vector field $\bm{\cal R}$, where $\tau$
is a continuous parameter specified along each curve ($x^{i}={\rm
const}$) of this congruence.

The congruence (\ref{th.8}) is the starting point of our
consideration. In general the quantities $x^{i}$ are not the first
four coordinates of a 5-dimensional coordinate system.

Hereinafter we will consider the 4-dimensional hypersurface
$\tau(y^{\nu}) = {\rm const}$ to be the 4-dimensional space-time.
Moreover, the parameter $\tau$ should be chosen to make tangent
vectors
\begin{eqnarray} \label{th.9}
         \xi^{\nu}(x^i, \tau) \equiv\>  \frac{\partial}{\partial \tau}
          y^{\nu}(x^i, \tau)
\end{eqnarray}
coinciding with the vectors $X^{\nu}$:
 \begin{eqnarray} \label{a.11}
 \xi^{\nu}(x^i, \tau) = X^{\nu}.
\end{eqnarray}
It is important to point out that the equation (\ref{a.11}) is only
valid in the frame $\bigl\{X^{\nu}\bigr\}$; but it can always be
rewritten in an arbitrary frame $\{y^{\nu}\}$: $ \xi^{\nu}
(x^{i},\tau) = {\cal R}^{\nu}$, where ${\cal R}^{\nu}$ are
components of the vector $\bm{\cal R}$ in the coordinate basis $
\bm{e}_{\nu}=\pderiv{}{y^{\nu}}$. Let us emphasize that all equations
containing vectors $X^\mu$ are only valid in the special frame.
Henceforth it will not be specially accentuated.

According to (\ref{2.4}) we postulate equality to zero of the Lie
derivative with respect to $\bm{{\cal R}}$ for any 4-dimensional
quantity (i.e. quantity which depends only on 4-dimensional
coordinates). It is quite natural to extend the introduced
postulate on all 5-dimensional vectors and tensors which further
will be associated with the 4-dimensional quantities :
\begin{eqnarray} \label{a.2}
        \Li{\bm{\cal R}}{\bm{T}}=0\qquad
        (\Li{\bm{\cal R}}{T^{\alpha\dots}\,_{\beta\dots}}=0).
\end{eqnarray}
In the coordinate basis (\ref{2.3}) one can rewrite the last
equation (\ref{a.2}) in the form
\begin{eqnarray} \label{g.6}
         T^{\mu_{1}\dots\mu_{n}}\,_{\nu_{1}\dots\nu_{m},\lambda}
         X^{\lambda} =
         (n-m)T^{\mu_{1}\dots\mu_{n}}\,_{\nu_{1}\dots\nu_{m}}.
\end{eqnarray}
The geometrical quantities satisfying the projector condition
(\ref{g.6}) are called projectors \cite{Dantzig,Veblen}. By
applying the projector condition (\ref{a.2}) to a metric tensor
\bm{g} we obtain
\begin{eqnarray} \label{g.7}
         \Li{\bm{\cal R}}{\bm{g}} = 0 \qquad
         (g_{\mu\nu , \varepsilon} X^{\varepsilon}
         = - 2g_{\mu\nu}).
\end{eqnarray}
From the last equation follows that the 5-dimensional radius
vector \bm{\cal R} is a Killing vector. Thus the congruence
(\ref{th.8}) is a Killing congruence (see for example \cite{Hawking}).

In order to study geometrical properties of this congruence we
introduce a unit vector \bm{s}, i.e.
\begin{eqnarray} \label{g.13}
          \bm{s}=\frac{\bm{\cal R}}{S}
          \qquad  \left( s^{\mu} = \frac{X^{\mu}}{S}\right),
\end{eqnarray}
where
     $\ds  S = \sqrt{\bm{g}(\bm{\cal R},\bm{\cal R})} =
          \sqrt{g_{\mu\nu}X^{\mu}X^{\nu}}$.
From the definition (\ref{g.13}) it is clear that \bm{s} is the
unit tangential vector field to the lines of the congruence
(\ref{th.8}).

In order to provide a description of geometrical properties of the
congruence (\ref{th.8}) we introduce, as usually, the following
quantities:
\begin{eqnarray} \label{g.15}
     a)\     G^{\mu}\equiv\>  s^{\nu}s^{\mu}\,_{;\nu},\quad
     b)\     \omega_{\mu\nu}\equiv\>
             P^{\tau}\,_{\mu}P^{\varepsilon}\,_{\nu}s_{[\tau,\varepsilon]}  \quad
     c)\     D_{\mu\nu}\equiv\>
             P^{\tau}\,_{\mu}P^{\varepsilon}\,_{\nu}s_{(\tau;\varepsilon)},
\end{eqnarray}
where
\begin{center}
\parbox[t]{0.9\textwidth}{
\begin{tabular}{llll}
$G^{\mu}$\hspace{3cm}&
$\Longrightarrow$&\hspace{3cm}&
\parbox{0.35\textwidth}{
the first curvature vector of the lines of the congruence;
}\\[4ex]
$\omega_{\mu\nu}$\hspace{3cm}&
$\Longrightarrow$&\hspace{3cm}&
\parbox{0.35\textwidth}{
 the angular velocity tensor  of the congruence;
}\\[4ex]
$D_{\mu\nu}$\hspace{3cm}&
$\Longrightarrow$&\hspace{3cm}&
\parbox{0.35\textwidth}{
rate-of-strain tensor  of the congruence.}
\\
\end{tabular}
}
\end{center}
The quantity
\begin{eqnarray} \label{g.16}
    P^{\tau}\,_{\mu} = \delta^{\tau}\,_{\mu} - s^{\tau}s_{\mu}
\end{eqnarray}
is the projection tensor. The semicolon means the Riemannian
covariant derivative ($\stackrel{R}{\nabla}$):
\begin{eqnarray} \label{g.18}
          \stackrel{R}{\nabla}_{\scriptstyle{\bm{e}}_{\tau}}\bm{e}_{\alpha} =
          \CR{\varepsilon}{\alpha\tau}\bm{e}_{\varepsilon}\quad
          \left( \bm{e}_{\alpha}=\pderiv{}{x^{\alpha}}\right)
\end{eqnarray}
with $\ds\CR{\varepsilon}{\alpha\tau}                \equiv\>
    \frac{1}{2}g^{\varepsilon\sigma}\left(g_{\sigma\alpha ,\tau}+
    g_{\tau\sigma  ,\alpha}  -  g_{\alpha\tau  ,\sigma}\right)$.
If we take into account that the vector field \bm{\cal R} is Killingian, we obtain
\begin{eqnarray}    \label{g.23}
\ds\abc{a} & G^{\mu} & =  \frac{1}{2S}
    X^{\varepsilon \mu}s_{\varepsilon }=
    - \frac{S^{,\mu}}{S}, \nonumber\\ \bigskip \ds
\abc{b}  &
    \omega_{\mu\nu} & =
    \frac{1}{2S}
    P^{\varepsilon}\,_{\mu}P^{\tau}\,_{\nu}X_{\tau\varepsilon},
 \nonumber\\ \bigskip \ds
\abc{c}& D_{\mu\nu} & =0,
    \end{eqnarray}
where the following abbreviation was used:
\begin{eqnarray} \label{g.25}
X_{\mu\nu} = X_{\nu ,\mu} - X_{\mu ,\nu}.
\end{eqnarray}
From the equations  (\ref{g.15}) and (\ref{g.23}) we obtain the
following important relations:
\begin{eqnarray} \label{k.56}
          s_{\mu;\nu} = D_{\mu\nu} + \omega_{\mu\nu} +
          G_{\mu}s_{\nu},
\end{eqnarray}
\begin{eqnarray} \label{k.59}
          \frac{1}{2S}X_{\nu\mu} = \omega_{\mu\nu} +
          \frac{1}{S}(s_{\mu}S_{,\nu} - s_{\nu}S_{,\mu}).
\end{eqnarray}

From the relation (\ref{g.23}b) follows that in general a holonomic
hypersurface orthogonal to the given congruence does not exist.
(The case $\omega_{\mu\nu} = 0$ is physically not interesting,
since further the angular velocity of the congruence will be
associated with the electromagnetic field). Therefore in contrast
to Schmutzer's orthogonality approach of space-time (based on the
basis vector formalism) here we want to offer an alternative
version of this problem: we shall identify space-time with a
4-dimensional hypersurface in the 5-dimensional space abandoning
the requirement of
\textbf{orthogonality} of this hypersurface to the congruence.

Let us consider some hypersurface $\tau(X^{\alpha}) = {\rm const}$.
\label{Hyper} As far as a parameter $\tau$ cannot univalently be
derived from the equations (\ref{th.9}) and (\ref{a.11}), then
hypersurfaces $\tau(X^{\alpha}) = {\rm const}$ are not defined
univalently either. Therefore we can choose in $\mathscr{M}_{5}$ an
arbitrary hypersurface which we shall identify with hypersurface
$\tau(X^{\alpha}) =0$. This hypersurface should only satisfy the
condition $X^{\alpha}\tau_{,\alpha}\not=0$. With an exponential map
we can extend it along the lines of congruence (\ref{th.8}) to a
finite region in $\mathscr{M}_{5}$. Thus we receive a
one-parametric set of hypersurfaces. Hence from the equations
(\ref{th.9}) and (\ref{a.11}) follows that
\begin{eqnarray} \label{g.28}
          <{\rm d}\tau , \bm{ \xi}> = 1 \qquad (
          \xi^{\varepsilon }\tau_{,\varepsilon } =
          X^{\varepsilon }\tau_{,\varepsilon } = 1)
\end{eqnarray}
and
\begin{eqnarray} \label{g.29}
          \Li{\bm{\cal R}}{{\rm d}\tau} = 0
           \qquad
         ( X^{\varepsilon }\tau_{,\alpha,\varepsilon }+
          X^{\varepsilon }\,_{,\alpha}\tau_{,\varepsilon } = 0).
\end{eqnarray}
From the last relation we can conclude that the one-form ${\rm
d}\tau $ which further we also shall denote by $\bm{\zeta}$
satisfies the projector condition (\ref{g.6}):
\begin{eqnarray} \label{g.30}
         \Li{\bm{\cal R}}{\bm{\zeta}} = 0,
                        \qquad \bm{\zeta} \equiv\>  {\rm d}\tau \qquad (\zeta_{\mu,\tau}
          X^{\tau} = - \zeta_{\mu}).
\end{eqnarray}
The unit one-form $\bm{\nu} = \Lambda \bm{\zeta}$ also fulfills this
condition:
\begin{eqnarray} \label{g.31}
          \Li{\bm{\cal R}}{\bm{\nu}} = 0 \qquad
          (\nu_{\tau,\mu}X^{\mu} = - \nu_{\tau}),
\end{eqnarray}
where
      $    \Lambda = <\bm{\nu},\bm{\cal R}> =
          \nu_{\varepsilon }X^{\varepsilon }$ and $
          \nu_{\varepsilon }\nu^{\varepsilon }=1.
      $

Above we introduced the projection tensor
$P^{\alpha}\,_{\varepsilon }$. However, the hypersurface
$\tau(X^{\alpha}) = 0$ (we also shall denote it by $\mathscr{M}_4$)
is not orthogonal to the congruence (\ref{th.8}). Therefore it is
possible to define two more projection tensors:
\begin{eqnarray} \label{g.33}
          \abc{a} b_{\alpha\varepsilon } \equiv\>  g_{\alpha\varepsilon }
          - \nu_{\alpha}\nu_{\varepsilon },\quad
          \abc{b} h^{\alpha}\,_{\varepsilon } \equiv\>
          g^{\alpha}\,_{\varepsilon } - \xi^{\alpha}\zeta_{\varepsilon }.
\end{eqnarray}
All these projection tensors satisfy the projector condition
(\ref{g.6}):
\begin{eqnarray} \label{g.34}
          P^{\alpha}\,_{\varepsilon ,\nu}X^{\nu} = 0,\quad
          h^{\alpha}\,_{\varepsilon ,\nu}X^{\nu} = 0,\quad
          b_{\alpha\varepsilon ,\nu}X^{\nu} = - 2 b_{\alpha\varepsilon }.
\end{eqnarray}
The projection tensor $b_{\alpha\varepsilon }$ sometimes is called
the first fundamental form of $\mathscr{ M}_{4}$ or the induced
metric on $\mathscr{ M}_{4}$. (In the following we shall define the
induced metric on $\mathscr{ M}_{4}$ in a somewhat different way).
The tensor $
\chi_{\alpha\varepsilon}$ defined on the hypersurface $\tau = 0$ by
\begin{eqnarray}\label{g.36}
        \chi_{\alpha\varepsilon}\equiv\>
    b^{\mu}\,_{\alpha}b^{\nu}\,_{\varepsilon  }   \nu_{(\mu;\nu)}    =
    \frac{1}{2\Lambda}\Li{\bm{\lambda}}b_{\alpha\varepsilon         },
\end{eqnarray}
is called the second fundamental form or the exterior curvature of
$\tau = 0$. Here the following abbreviations were used:
\begin{eqnarray}\label{g.37}
    \lambda^{\varepsilon } \equiv\> \Lambda\nu^{\varepsilon }
          = X^{\varepsilon }- \mathscr{ X}^{\varepsilon },\quad
   \mathscr{X}^{\varepsilon }\equiv\>  b^{\varepsilon
   }\,_{\alpha}X^{\alpha}.
\end{eqnarray}

The above introduced projection tensors in general differ from each
other. Therefore the question, which of them should be used for the
projection of 5-dimensional vectors and tensors into the
4-dimensional hypersurface, is not trivial. In order to give an
answer to this question, we consider the map $\phi$:
\begin{eqnarray} \label{th.10}
     \phi:\quad\mathscr{M}_5 \;
     \stackrel{\phi}{\longrightarrow} \; \mathscr{ M}_4.
\end{eqnarray}

The map $\phi$ should be defined in such a way to make mapped
quantities not depending on the parameter $\tau$, i.e. on the
``fifth coordinate'' (cylinder condition). This requirement means
that all points laying on the same line of the congruence are
mapped to the same point on the hypersurface $\tau(X^{\nu}) = {\rm
const}$. The elementary map of this type is an exponential map (see
Fig. 1).

 \begin{figure}[hbt]  \label{abb}
                        \begin{center}
                 \usebox1 %
 \caption{\textsf{On the introduction of the  exponential map.}
         $P^{\prime} = \phi_{\tau_1}(P_{1}) =
\phi_{\tau_2}(P_{2}).\
P^{\prime} \in \mathscr{ M}_{4}$
($\mathscr{ M}_{4}:\; \tau(X^{\alpha})=0$)}.
\end{center}
\end{figure}

The coordinates of the point $P_{1}$ satisfy the relation
\begin{eqnarray}\label{a.43}
          X^{\alpha}(P_{1})=X^{\alpha}(x^{m}_{0},\tau_{1})=
          X^{\alpha}(x^{m}_{0},0)\exp(\tau_{1}),
\end{eqnarray}
where $X^{\alpha}(x^{m}_{0},0) =X^{\alpha}(P^{'})$. Using the
equations (\ref{th.9}) and (\ref{a.11}) one can obtain
\begin{eqnarray} \label{a.44}
          \phi_{\tau}: \quad
          X^{\alpha}(P) = \exp(\tau)X^{\alpha}(\phi_{\tau}(P)).
\end{eqnarray}
Now we have to discuss how the vectors and tensors are transformed
by the map $\phi_{\tau}$.

Let \bm{V} be a tangent vector to the curve $\lambda(t)$ at the
point $P_{1}$, having the following form in local coordinates in a
neighborhood of the point $P_{1}$:
\begin{eqnarray} \label{a.45}
          X^{\alpha}(\lambda(t)) = X^{\alpha}(P_{1}) +
          t V^{\alpha},
\end{eqnarray}
where $V^{\alpha}$ are the components of the vector \bm{V}
($\bm{V}= \pderiv{}{t}$) in the coordinate basis $\bm{E}_{\alpha}$,
i.e.
\begin{eqnarray} \label{a.46}
          \bm{V} = V^{\alpha} \pderiv{}{X^{\alpha}}.
\end{eqnarray}
Comparing the equation (\ref{a.45}) with the following series
expansion:
\begin{eqnarray} \label{a.47}
          X^{\alpha}(\lambda(t)) &=&
          X^{\alpha}\bigl(x^{m}(\lambda(t)),\tau(\lambda(t))\bigr)
        =  \nonumber\\ \bigskip \ds
          &=& X^{\alpha}(x^{m}_{0},\tau_{1}) +
    \left.
          \left( X^{\alpha}\,_{,m}\deriv{x^{m}}{t} +
          \pderiv{X^{\alpha}}{\tau}\deriv{\tau}{t}\right)
    \right|_{P_{1}}
          \cdot t  \nonumber\\ \bigskip \ds
          & + & O(t^{2}),
\end{eqnarray}
we obtain
 \begin{eqnarray} \label{a.48}
        \left.  V^{\alpha}\right|_{P_{1}} =
          \left(X^{\alpha}\,_{,m}\deriv{x^{m}}{t} +
          \xi^{\alpha} \left.\deriv{\tau}{t}  \right)\right|_{P_{1}}
          \quad \left(
          \xi^{\alpha} = \left.\deriv{X^{\alpha}}{\tau}
 \right|_{x^{m}={\rm const}}\right).
\end{eqnarray}
The curve $\lambda(t)$ can be projected by the exponential map
$\phi_{\tau(\lambda)}$ onto the hypersurface $\tau=0$. The notation
$\phi_{\tau(\lambda)}$ should accentuate that each point of the
curve $\lambda(t)$ is mapped by the proper exponential map
$\phi_{\tau}$ ($\tau$ depends on $t$ ). We denote the mapped curve
obtained by this procedure by $\gamma(t)$:
\begin{eqnarray} \label{a.49}
    \phi(\lambda(t)) =  \gamma(t),
\end{eqnarray}
where $\phi$ means $\phi_{\tau(\lambda(t))}$.

Further we shall consider only vector fields \bm{V} commuting with
$\pderiv{}{\tau}$, i.e. the vector fields being projectors. In this
case the maps of curves $\lambda(t)$ and $\lambda^{'}(t)$
($\lambda^{'}(0) =
    P^{'}$) coincide:
\begin{eqnarray} \label{a.50}
     \phi(\lambda(t)) =  \phi(\lambda^{'}(t)) = \gamma(t),
\end{eqnarray}
where $ \lambda^{'}(t) = \phi_{\tau_{1}}(\lambda(t))$. Therefore,
without any further restriction we may consider only such curves
whose initial points $P_{1}$ ($P_{1} = \lambda(0)$) belong to the
hypersurface $\tau = 0$, i.e. $P_{1}=P^{'} \in \mathscr{ M}_{4}$.

For the  mapped curve $\gamma(t)$ following expansion is valid
\begin{eqnarray} \label{a.51}
          X^{\alpha}(\gamma(t)) &=&
          X^{\alpha}\left[\left(x^{m}_{0} + \deriv{x^{m}}{t} +
          O(t^{2})\right),0\right] \nonumber\\ \bigskip \ds
&=&          X^{\alpha}(x^{m}_{0},0) +
          \left. \left(X^{\alpha}\,_{,m}\deriv{x^{m}}{t}
          \right)\right|_{P} \cdot t + O(t^{2}).
\end{eqnarray}
From the equation (\ref{a.48}) follows
\begin{eqnarray} \label{a.51a}
          V^{\alpha} = \left(
          \deriv{x^{m}}{t}\bm{e}_{m} + \deriv{\tau}{t}\bm{\xi}
          \right)(X^{\alpha}),
\end{eqnarray}
where the vectors are defined by
\begin{eqnarray} \label{a.52}
          \bm{e}_{m} = \pderiv{}{x^{m}},\quad
          \bm{\xi} = \pderiv{}{\tau}.
\end{eqnarray}
The equation (\ref{a.51a}) implies that the following relation for
the vector field \bm{V} is fulfilled:
\begin{eqnarray} \label{a.53}
          \bm{V}=\deriv{x^{m}}{t}\bm{e}_{m} +
          \deriv{\tau}{t}\bm{\xi}.
\end{eqnarray}
Thus at the point $P_{1}$ ($P_{1} \in \mathscr{ M}_{4}; \gamma(0)
=\lambda^{'}(0) = P_{1} $) the 4-dimensional $T_{P_{1}}$ and
5-dimensional $T_{\phi(P_{1})}$ vector spaces can be constructed
as follows \cite{Hawking}:
\begin{eqnarray} \label{g.49}
     \bm{V} \in T_{P_{1}}, \qquad
     \phi_{*}\bm{V}\in T_{\phi(P_{1})}.
\end{eqnarray}
Here we used the abbreviations (compare with \cite{Hawking}):
\begin{eqnarray} \label{g.50}
    \phi_{*}\bm{V} \equiv\>  \left. \left(
    \pderiv{}{t}\right)_{\gamma} \right|_{\phi(P_{1})} =
    \left. \deriv{x^{m}}{t}\right|_{P_{1}}\bm{e}_{m},
\end{eqnarray}
where
\begin{eqnarray} \label{g.51}
    \deriv{x^{m}}{t} = x^{m}\,_{,\alpha}V^{\alpha}.
\end{eqnarray}
It is necessary to note that $ \phi(P_{1}) = P_{1}$, $\gamma(0) =
    \lambda^{'}(0) = P_{1} $.

It is easy to show that the one-form
$\bm{e}^{m}\equiv\> {\rm d}x^{m}$ and the vectors $\bm{e}_{m}    =
    \pderiv{}{x^{m}}$ satisfy the equations
\begin{eqnarray} \label{g.52}
          <\bm{\zeta}, \bm{e}_{m}> = 0, \qquad
          <\bm{e}^{m},\bm{\xi}> = 0.
\end{eqnarray}
These equations imply that one can rewrite the projector
$h^{\alpha}\,_{\varepsilon }$ in the form
\begin{eqnarray} \label{g.53}
          h^{\alpha}\,_{\varepsilon } = g^{\alpha}\,_{m}
          g^{m}\,_{\varepsilon } = g^{\alpha}\,_{\varepsilon } -
          \xi^{\alpha}\zeta_{\varepsilon },
\end{eqnarray}
where we used the definitions \cite{Schmutzer1}
\begin{eqnarray} \label{g.54}
          \abc{a} g^{m}\,_{\varepsilon } =<\bm{e}^{m},\bm{e}_{\varepsilon }>
          = x^{m}\,_{,\varepsilon },\qquad
          \abc{b}
          g^{\varepsilon }\,_{m} =<\bm{e}^{\varepsilon },\bm{e}_{m}>
          = X^{\varepsilon }\,_{,m}.
\end{eqnarray}
Apart from that it is easy to show that between the quantities
$\bm{e}_{m}$, $\bm{e}^{m}$,
    $\bm{e}_{\varepsilon  }$
and
$\bm{e}^{\varepsilon  }$
the following relation is valid:
\begin{eqnarray} \label{g.56}
          \abc{a}
          \bm{e}_{\varepsilon } = g^{m}\,_{\varepsilon }\bm{e}_{m}+
          \zeta_{\varepsilon }\bm{\xi}, \quad
          \abc{b}
          \bm{e}^{\varepsilon } = g^{\varepsilon }\,_{m}{\rm d}x^{m}+
          \xi^{\varepsilon }{\rm d}\tau.
\end{eqnarray}
The last relation and the definition (\ref{g.50}) lead us to
\begin{eqnarray} \label{g.56a}
           \til{\bm{V}} \equiv\>  \phi_{*}\bm{V} =
           (g^{m}\,_{\alpha}V^{\alpha})\bm{e}_{m} =
           \til{V}^{\varepsilon }\bm{e}_{\varepsilon },
\end{eqnarray}
where we used the abbreviation
\begin{eqnarray} \label{g.57}
          \til{V}^{\varepsilon } \equiv\>
          h^{\varepsilon }\,_{\alpha}V^{\alpha}.
\end{eqnarray}
Thus in the tangent vector space $T_{P}$ it is possible to define a
4-dimensional subspace $\til{T}_{P_{1}}$ ($\til{T}_{P_{1}} \subset
T_{P_{1}} $):
\begin{eqnarray} \label{g.58}
    \til{T}_{P} = \bigl\{
    \til{\bm{V}}: \; \forall \til{\bm{V}}\in \til{T}_{P},
    \quad
    \phi_{*}\til{\bm{V}} = \til{\bm{V}}
\bigr\}.
\end{eqnarray}
The equation
$\phi_{*}\bm{V}=\til{V}^{\varepsilon }\bm{e}_{\varepsilon    }$
should be interpreted in the following way:
\begin{eqnarray}\label{AA1}
  \phi_{*}\bm{V} = (x^{m}\,_{,\alpha}V^{\alpha})\bm{e}_{m}
  \in T(\mathscr{M}_{4}).
\end{eqnarray}
In a vector space $T(\mathscr{M}_{5})$ it is possible to construct
a 4-dimensional vector space formed by the vectors of the type
 $(h^{\varepsilon }\,_{\alpha}V^{\alpha})\bm{e}_{\varepsilon }$.
The spaces $T(\mathscr{ M}_{4})$ and $\til{T}(\mathscr{M}_{5})$ are
isomorphic:
\begin{eqnarray}\label{AA2}
  (x^{m}\,_{,\alpha}V^{\alpha})\bm{e}_{m}
  \Longleftrightarrow
  (h^{\varepsilon
    }\,_{\alpha}V^{\alpha})\bm{e}_{\varepsilon }.
\end{eqnarray}

The map $\phi_{*}$,  namely
$$
    T_{P}(\mathscr{M}_{5}) \stackrel{\phi_{*}}{\longrightarrow}
    T_{\phi(P)}(\mathscr{ M}_{4}),
$$
naturally induces the map $\phi^{*}$ for the one-forms:
$$
        T^{*}_{\phi(P)}(\mathscr{ M}_{4}) \stackrel{\phi^{*}}{\longrightarrow}
        T^{*}_{P}(\mathscr{M}_{5}),
$$
where for all $ \bm{\omega}\in T^{*}_{\phi(P_{1})}$ and for all
$\bm{V}\in T_{P_{1}}$ the next relation is valid:
\begin{eqnarray} \label{g.59}
          \left.<\phi^{*}\bm{\omega},\bm{V}>\right|_{P_{1}} =
          \left.<\bm{\omega},\phi_{*}\bm{V}>\right|_{\phi(P_{1})}.
 \end{eqnarray}
The set of all one-forms satisfying the relation
\begin{eqnarray} \label{g.60}
    \phi^{*}\til{\bm{\omega}} = \til{\bm{\omega}}
\end{eqnarray}
forms a linear space $\til{T}^{*}(\mathscr{
    M}_{5}) \subset  T^{*}(\mathscr{  M}_{5})$,
where $T^{*}(\mathscr{ M}_{5})$ is the space of all one-forms. From
(\ref{g.59}) follows that for any one-form from
$\til{T}^{*}(\mathscr{M}_{5})$ holds
\begin{eqnarray} \label{g.61}
          <\til{\bm{\omega}}, \bm{\xi}> = 0,\qquad
          (\quad \til{\omega}_{\alpha}X^{\alpha} = 0).
\end{eqnarray}

There are two possible ways to associate elements of
$T^{*}(\mathscr{M}_{5})$ with elements of
$\til{T}^{*}(\mathscr{M}_{5})$:
$$
\abc{a}    \omega_{\alpha} \longrightarrow \til{\omega}_{\alpha}
    = h^{\varepsilon }\,_{\alpha}\omega_{\varepsilon },
$$
$$
\abc{b}    \omega_{\alpha} \longrightarrow \til{\omega}_{\alpha}
    = P^{\varepsilon }\,_{\alpha}\omega_{\varepsilon }.
$$
In both cases the quantities $\til{\omega}_{\alpha}$ satisfy the
equation (\ref{g.61}) automatically. Hereinafter quantities with
tilde  will be associated with physical quantities in the
space-time. However, the definition a) cannot be accepted, as in
this case the following relations would be valid:
$$
     \til{V}^{\alpha} \equiv\>  h^{\alpha}\,_{\varepsilon }V^{\varepsilon } \not=
     \til{g}^{\alpha\varepsilon }\til{M}_{\varepsilon }, \quad
     \til{M}_{\alpha} \equiv\>  h^{\varepsilon }\,_{\alpha }V_{\varepsilon } \not=
     \til{g}_{\varepsilon \alpha }\til{V}^{\varepsilon },
$$
where
$$
     \til{g}_{\alpha\beta} =
     h^{\varepsilon }\,_{\alpha}h^{\sigma}\,_{\beta}g_{\varepsilon \sigma},
\quad     \til{g}^{\alpha\sigma} =
     h^{\alpha}\,_{\nu}h^{\sigma}\,_{\mu}g^{\nu \mu}.
$$
On the contrary, the definition b) is consistent. In this case the
following relations will be valid:
\begin{eqnarray} \label{g.62}
\abc{a}
    \til{g}_{\alpha\beta} & \equiv\>  &
     P^{\varepsilon }\,_{\alpha}P^{\sigma}\,_{\beta}g_{\varepsilon  \sigma} =
    P_{\alpha\beta}, \nonumber\\
\abc{b}
        \til{g}^{\alpha\sigma} & \equiv\>  &
        h^{\alpha}\,_{\nu}h^{\sigma}\,_{\mu}g^{\nu \mu} =
        g^{\alpha\sigma} - 2 X^{(\alpha}\zeta^{\sigma)}+
        \frac{1}{\Lambda^{2}}X^{\alpha}X^{\sigma},  \nonumber\\
\abc{c}
       \til{g}^{\mu}\,_{\nu} & \equiv\>  &
       h^{\mu}\,_{\varepsilon }P^{\alpha}\,_{\nu}g^{\varepsilon }_{\alpha}
     = h^{\mu}\,_{\nu}.
    \end{eqnarray}
 Using (\ref{g.62}), for an arbitrary vector \bm{V} we obtain:
\begin{eqnarray} \label{g.64}
    \til{V}^{\alpha} = \til{g}^{\alpha\varepsilon}\til{V}_{\varepsilon },
    \qquad
    \til{V}_{\alpha} = \til{g}_{\alpha\varepsilon}\til{V}^{\varepsilon
    }.
\end{eqnarray}

The last results can be summarized in the sentence: The
5-dimensional tensors are to be projected onto hypersurface
$\tau(X^{\alpha}) = 0$ (projected quantities are denoted by a
tilde) with the help of the procedure:
\begin{eqnarray} \label{th.11}
     T^{{\mu}\cdots}\,_{\cdots {\nu}} \; \stackrel{\phi}{\longrightarrow} \;
     \tilde{T}^{{\mu}\cdots}\,_{\cdots {\nu}} \equiv\>
     h^{\mu}\,_{\sigma}\cdots P^{\tau}\,_{\nu}\cdots T^{\sigma\cdots}\,_{
     \cdots\tau}.
\end{eqnarray}

The quantities $x^{i}$ being introduced as parameters earlier and
parametrizing the congruence (\ref{th.8}) can be used as
coordinates in $\mathscr{ M}_{4}$. Let us point out that it is
necessary to require a certain continuity for the quantities
$x^{i}$. Apart from that these quantities are defined accurately
within the following transformation: $x^{i}\longrightarrow x^{i'} =
x^{i'}(x^{j})$. In this case the vectors $\bm{e}_{m}=
\pderiv{}{x^{m}}$ and the one-forms $\bm{e}^{m}={\rm d}x^{m}$ form
a basis in $T(\mathscr{M}_{4})$ and $T^{*}(\mathscr{ M}_{4})$,
either. These bases satisfy the following relations:
\begin{eqnarray} \label{g.66}
          <\bm{e}^{m},\bm{e}_{n}> = \delta^{m}\,_{n}, \quad
          \bigl[\bm{\cal R},\bm{e}_{m}\bigr] = 0, \quad
          <\bm{\nu},\bm{e}_{m}> = 0.
\end{eqnarray}
Using the equations (\ref{g.37}), (\ref{g.54}) and (\ref{g.56}), we
can find several important relations:
\begin{eqnarray} \label{g.67}
                \abc{a}
    \bm{e}_{\alpha}h^{\alpha}\,_{\varepsilon } =
    g^{m}\,_{\varepsilon }\bm{e}_{m}, \quad
\abc{b}
       \bm{e}^{\alpha}h^{\varepsilon }\,_{\alpha } =
    g^{\varepsilon }\,_{m}\bm{e}^{m};
\end{eqnarray}
\begin{eqnarray} \label{g.68}
                \abc{a}
    \bm{e}_{\alpha}P^{\alpha}\,_{\varepsilon } =
    g^{m}\,_{\varepsilon }\bm{e}_{m} +
(\zeta_{\varepsilon } - s_{\varepsilon })\bm{s},
 \quad
\abc{b}
       \bm{e}^{\alpha}P^{\varepsilon }\,_{\alpha } =
    (g^{\alpha}\,_{m}P^{\varepsilon}\,_{\alpha})\bm{e}^{m};
\end{eqnarray}
\begin{eqnarray} \label{g.69}
                \abc{a}
    \bm{e}_{\alpha}b^{\alpha}\,_{\varepsilon } =
    (g^{m}\,_{\alpha }b^{\alpha}\,_{\varepsilon })\bm{e}_{m}, \quad
                      \abc{b}
       \bm{e}^{\alpha}b^{\varepsilon }\,_{\alpha } =
    g^{\varepsilon }\,_{m}\bm{e}^{m}+ \mathscr{X}^{\varepsilon }{\rm d}\tau .
\end{eqnarray}
We already mentioned that the tangent spaces $T(\mathscr{
M}_{4})$ and $\til{T}(\mathscr{ M}_{5})$ are isomorphic.
Therefore one can write:
\begin{eqnarray} \label{g.70}
T^{m} \equiv\>  g^{m}\,_{\varepsilon }\til{T}^{\varepsilon }, \qquad
\til{T}^{\varepsilon } = g^{\varepsilon }\,_{m}T^{m},
\end{eqnarray}
\begin{eqnarray} \label{g.71}
          \omega_{m} \equiv\>  g^{\varepsilon }\,_{m}\til{\omega},
          \quad
          \til{\omega}_{\varepsilon } = g^{m}\,_{\varepsilon }\omega_{m}.
\end{eqnarray}
Thus the projection procedure from $T(\mathscr{M}_{5})$ into
$T(\mathscr{ M}_{4})$ is defined as follows:
\begin{eqnarray} \label{g.72}
          T^{\alpha} \longrightarrow T^{m} = \til{g}^{m}\,_{\varepsilon }
          T^{\varepsilon }, \quad
          \omega _{\alpha} \longrightarrow
          \omega_{m} = \til{g}^{\varepsilon }\,_{m}\omega_{\varepsilon },
\end{eqnarray}
where we used the abbreviation
\begin{eqnarray} \label{g.73}
\til{g}^{\varepsilon }\,_{m} \equiv\>  P^{\varepsilon }\,_{\alpha}
g^{\alpha}\,_{m} \not= g^{\varepsilon}\,_{m}, \quad
\til{g}^{m}\,_{\varepsilon } \equiv\>  h^{\alpha}\,_{\varepsilon }
g^{m}\,_{\alpha} = g^{m}\,_{\varepsilon }.
\end{eqnarray}

The metric induced on the hypersurface $\mathscr{ M}_{4}$ will be
denoted further by $\til{\bm{g}}$ (in the theory of surfaces one
understands under the induced metric the quantity $b_{\mu\nu}$
defined by means of (\ref{g.33}a) \cite{Hawking}). This metric
satisfies the following relations:
\begin{eqnarray} \label{g.74}\abc{a}
     \til{g}_{mn} \equiv\>  \til{g}^{\alpha}\,_{m}
     \til{g}^{\beta}\,_{n}g_{\alpha\beta},\quad
\abc{b}
\til{g}^{mn} \equiv\>  \til{g}^{m}\,_{\alpha}
     \til{g}^{n}\,_{\beta}g^{\alpha\beta}, \nonumber \\
\abc{c}
\til{g}^{m}\,_{n} \equiv\>  \til{g}^{m}\,_{\alpha}
     \til{g}^{\varepsilon }\,_{n}g^{\alpha}\,_{\varepsilon} =
     \delta^{m}\,_{n},\quad
\abc{d}
\til{g}^{mn}\til{g}_{mk} =  \delta^{n}\,_{k}.
\end{eqnarray}
It is necessary to accentuate that the relation
\begin{eqnarray} \label{g.75}
\bm{g}(\bm{e}_{m},\bm{e}_{n})  \not= \til{g}_{mn},\quad
\til{g}_{mn} = \til{\bm{g}}(\bm{e}_{m},\bm{e}_{n})
\end{eqnarray}
is fulfilled, where
\begin{eqnarray} \label{g.76}
          \bm{g} \equiv\>
          g_{\alpha\varepsilon }
          {\rm d}X^{\alpha}\otimes{\rm d}X^{\varepsilon },
\quad
\til{\bm{g}} \equiv\>  \til{g}_{mn}
{\rm d}x^{m}\otimes{\rm d}x^{n}.
\end{eqnarray}
At the end we have to present further two important relations that
immediately follow from (\ref{g.66}):
\begin{eqnarray} \label{g.77}
\abc{a}
x^{m}\,_{,\alpha}X^{\alpha} = 0, \quad
\abc{b}
X^{\varepsilon }\,_{,m}\zeta_{\varepsilon } =0.
\end{eqnarray}

\section{Field Equations}

In the introduction we mentioned that in the course of four decades
three neighboring versions of Schmutzer's 5-dimensional Projective
Unified Field Theory came into being. All these versions are based
on the following 5-dimensional field equations:
\begin{eqnarray} \label{g.96}
    \mathscr{G}_{\alpha\varepsilon }
    = \kappa_{0}\theta_{\alpha\varepsilon }.
\end{eqnarray}
The explicit expression of the symmetric tensor $
\mathscr{G}_{\alpha\varepsilon }$ for the versions II and III of PUFT can
be found, using the equations (\ref{W1}) and (\ref{W17}),
respectively. In order to obtain a 4-dimensional field equations
these 5-dimensional equations have to be projected onto the
4-dimensional space-time. The equation (\ref{g.96}) can always be
written in the following form:
\begin{eqnarray} \label{i.7}
\til{\mathscr{G}}_{\mu\nu} + 2\til{\mathscr{G}}_{(\mu}s_{\nu )}+
\mathscr{G}s_{\mu}s_{\nu} =
\kappa_{0}(
\til{\stackrel{5}{\theta}}_{\mu\nu} +
2\til{\stackrel{5}{\theta}}_{(\mu}s_{\nu )}+
\til{\stackrel{5}{\theta}}s_{\mu}s_{\nu}),
\end{eqnarray}
where the  abbreviations are given by
\begin{eqnarray} \label{i.9}
\begin{array}{lll}  \ds \bigskip
    \abc{a}\til{\mathscr{G}}_{\mu\nu}
    \equiv\>  P^{\alpha}\,_{\mu}P^{\beta}\,_{\nu}\mathscr{G}_{\alpha\beta},
    \quad
    &\abc{b}
    \til{\mathscr{G}}_{\mu}
    \equiv\>  P^{\alpha}\,_{\mu}s^{\beta}\mathscr{G}_{\alpha\beta},
    \quad
    &\abc{c}
    \mathscr{G}
    \equiv\>  s^{\alpha}s^{\beta}\mathscr{G}_{\alpha\beta},\\ \ds
    \abc{d}\til{\stackrel{5}{\theta}}_{\mu\nu}
    \equiv\>  P^{\alpha}\,_{\mu}P^{\beta}\,_{\nu}{\theta}_{\alpha\beta},
    \quad
    &\abc{e}
    \til{\stackrel{5}{\theta}}_{\mu}
    \equiv\>  P^{\alpha}\,_{\mu}s^{\beta}{\theta}_{\alpha\beta},
    \quad
    &\abc{f}
    {\stackrel{5}{\theta}}
    \equiv\>  s^{\alpha}s^{\beta}{\theta}_{\alpha\beta}.\\
\end{array}
\end{eqnarray}
It is easy to see that the equation (\ref{i.7}) is equivalent to
the following set of equations:
\begin{eqnarray} \label{i.10}
\begin{array}{l}\ds    \bigskip
\abc{a}\quad
     \til{\mathscr{G}}_{\mu\nu} = \kappa_{0}
     \til{\stackrel{5}{\theta}}_{\mu\nu}, \\ \bigskip\ds
\abc{b}\quad
     \til{\mathscr{G}}_{\mu} = \kappa_{0}
     \til{\stackrel{5}{\theta}}_{\mu}, \\  \bigskip\ds
\abc{c}\quad
     \mathscr{G} = \kappa_{0}
     {\stackrel{5}{\theta}} \\
\end{array}
\end{eqnarray}
Further we can see that the following correspondence is valid:
\begin{center}
\parbox[t]{0.8\textwidth}{
\begin{tabular}{llll}
\hspace{0.7cm}&equation  (\ref{i.10}a)\hspace{1cm}&
$\Longleftrightarrow$\hspace{1cm}&
\parbox{0.35\textwidth}{
 generalized Einstein equations,
}\\[5ex]
& equation (\ref{i.10}b)\hspace{1cm}&
$\Longleftrightarrow$\hspace{1cm}&
\parbox{0.35\textwidth}{

generalized  Maxwell equations,
}\\[5ex]
& equation (\ref{i.10}c)\hspace{1cm}&
$\Longleftrightarrow$\hspace{1cm}&
\parbox{0.35\textwidth}{

Field equation of the scalaric field.
}\\
\end{tabular}
}
\end{center}
Henceforth the 4-dimensional hypersurface $\mathscr{ M}_{4}$ will
be identified with the space-time. The physical metrics
$\stackrel{4}{\bm{g}}$ of the space-time can be defined in
different ways (all space-time quantities will be denoted by an
index ``4''). For example, we can identify the 4-dimensional
physical metric $\stackrel{4}{\bm{g}}$ with the metric
$\til{\bm{g}}$ induced on the hypersurface $\mathscr{ M}_{4}$:
 \begin{eqnarray}\label{k.1a}
         {\bm{\ge}} = \til{\bm{g}}\quad
         \quad
         (\ge_{\mu\nu} = \til{g}_{\mu\nu},
         \quad
         \ge^{\mu\nu} = \til{g}^{\mu\nu}).
\end{eqnarray}
In this case we obtain version I or III of PUFT if we use the
5-dimensional equations (\ref{W1}) or (\ref{W17}), respectively.
However, it is physically possible to connect these metrics
$\stackrel{4}{\bm{g}}$ and $\til{\bm{g}}$ by a conformal
transformation:
\begin{eqnarray} \label{k.1}
         {\bm{\ge}} = e^{\sigma}\til{\bm{g}}\quad
         \quad
         (\ge_{\mu\nu} = e^{\sigma}\til{g}_{\mu\nu},
         \quad
         \ge^{\mu\nu} = e^{- \sigma}\til{g}^{\mu\nu}).
\end{eqnarray}
In this case the 5-dimensional Einstein-like equations (\ref{W1})
lead to the system of equations of version II of PUFT. In order to
consider both these cases simultaneously we rewrite the equations
(\ref{k.1a}) and (\ref{k.1}) in the form:
 \begin{eqnarray}\label{k.1b}
{\bm{\ge}} = e^{\epsilon \sigma}\til{\bm{g}}\quad
\mbox{with}\quad
\epsilon =\left\{
 \begin{array}{lcl}
  0&\longleftrightarrow &\mbox{\textsf{Version I}+\textsf{Version III}}\\
  1&\longleftrightarrow &\mbox{\textsf{Version II}}\\
  \end{array}
\right.
\end{eqnarray}
The projection formalism can be simplified by using a
non-Riemannian connection in the 5-dimensional space and
considering the Riemannian connection in the 4-dimensional
space-time as induced.

\subsection{Connection on $\mathscr{M}_{5}$}

Let us introduce an induced (affine) connection on the hypersurface
$\mathscr{ M}_{4}$ denoted hereinafter as \nabv{}. The induced
connection \nabv{} and the connection on $\mathscr{ M}_{5}$
(denoted as
\nabf{}) are connected in the following way:
\begin{eqnarray} \label{k.2}
        \nabv{\varepsilon} \til{T}^{\mu\dots}\,_{\nu\dots}=
        h^{\mu}\,_{\sigma}\dots P^{\alpha}\,_{\varepsilon}
        P^{\lambda}\,_{\nu}\dots \til{T}^{\sigma\dots}\,
        _{\lambda\dots\DD \alpha},
\end{eqnarray}
where
\begin{eqnarray} \label{g.79}
     T^{\dots}\,_{\dots\DD\alpha} = \nabf{\alpha}
     T^{\dots}\,_{\dots},\quad
     \nabf{\alpha}\equiv\> \nabf{\bm{e}_{\alpha}},\quad
     \nabv{\alpha}\equiv\> \nabv{\bm{e}_{\alpha}}.
\end{eqnarray}
Henceforth we assume that the connection on $\mathscr{
M}_{4}$ is Riemannian, i.e. metrical and symmetric:
\begin{eqnarray} \label{k.3}
       \abc{a}
       \nabv{\varepsilon }\stackrel{4}{g}_{\mu\nu} = 0,\quad
       \abc{b}
       \nabv{\alpha}\nabv{\varepsilon }f =
       \nabv{\varepsilon }\nabv{\alpha}f.
\end{eqnarray}
Since the 4-dimensional covariant derivative is defined only for
the projected vectors (see (\ref{k.2})), the function $f $ should
satisfy the condition (\ref{a.2}): $\Li{\bm{\cal R}}{f} =
f_{,\alpha}X^{\alpha} = 0$. The 4-dimensional covariant derivative
(with respect to \nabv{}) in the direction of the basis vectors
$\bm{e}_{m}$ ($\bm{e}_{m} = \pderiv{}{x^{m}}$) will be denoted by a
semicolon:
\begin{eqnarray} \label{i.1}
         \nabv{\bm{e}_{m}}\til{T}^{k}\equiv\>
         \nabv{m}\til{T}^{k}\equiv\>
         \til{T}^{k}\,_{;m} =
         g^{k}\,_{\alpha}g^{\varepsilon }\,_{m}\nabv{\varepsilon }\til{T}^{\alpha},
\end{eqnarray}
where
$         \til{\bm{T}} = \bm{e}_{m}\til{T}^{m} =
         \bm{e}_{\alpha}\til{T}^{\alpha}.
$

As it is well known, the Riemannian connection is completely
defined by means of a metric. Therefore the relations (\ref{k.3})
are in fact conditions for the 5-dimensional connection \nabf{}. In
particular, from (\ref{k.3}) follows that the 5-dimensional
connection \nabf{} has to satisfy the following relations:
\begin{eqnarray} \label{k.10}
\begin{array}{ll}\ds  \bigskip
\abc{a}&
       \til{Q}_{\varepsilon \alpha\beta}\equiv\>
       P^{\tau}\,_{\varepsilon }P^{\nu}\,_{\alpha}P^{\mu}\,_{\beta}
       Q_{\tau\nu\mu}
        = \epsilon \sigma_{,\varepsilon }
       \til{g}_{\alpha\beta} =\epsilon  \sigma_{,\varepsilon }
       P_{\alpha\beta}, \\ \bigskip \ds
\abc{b} &
       \til{S}_{\alpha\beta}\,^{\tau}\equiv\>
       P^{\nu}\,_{\alpha}P^{\mu}\,_{\beta}h^{\tau}\,_{\varepsilon }
       S_{\nu\mu}\,^{\varepsilon } =0, \\ \ds
\abc{c} &
       h^{\mu}\,_{\alpha}P^{\nu}\,_{\beta}X^{\alpha}\,_{\DD\nu}
       = 0,
\end{array}
\end{eqnarray}
where the usual definitions
\begin{eqnarray} \label{k.6}
\abc{a}  g_{\alpha\beta\DD\varepsilon } = - Q_{\varepsilon \alpha\beta },
         \quad \abc{b}
         g^{\alpha\beta}\,_{\DD\varepsilon }
                   = Q_{\varepsilon }\,^{\alpha\beta} \quad
\abc{c}
        S_{\alpha \beta }\,^{\gamma }  =
        \Gamma_{[\alpha \beta ]}\,^{\gamma }
\end{eqnarray}
are used. One can easily verify that the 5-dimensional connection
in general is nonsymmetric and nonmetrical.

For this reason we write the 5-dimensional connection coefficients
$\Gamma^{\varepsilon }\,_{\mu\nu}$ in the following form:
\begin{eqnarray} \label{k.23}
          \Gamma^{\varepsilon }\,_{\mu\nu} =
          \CR{\varepsilon }{\mu\nu} +
          \sigma_{\mu\nu}\,^{\varepsilon },
\end{eqnarray}
    where
\begin{eqnarray} \label{k.26}
       \sigma_{\mu\nu}\,^{\varepsilon } =
       - (S^{\varepsilon }\,_{\nu\mu} + S_{\nu\mu}\,^{\varepsilon } -
            S_{\mu}\,^{\varepsilon }\,_{\nu})
       +  \frac{1}{2}
            (Q_{\nu\mu}\,^{\varepsilon } + Q_{\mu}\,^{\varepsilon }\,_{\nu}
           - Q^{\varepsilon }\,_{\nu\mu}).
\end{eqnarray}
The 5-dimensional connection cannot be found uniquely from the
demands (\ref{k.10}). However, the 5-dimensional field equations
(see (\ref{W1}) and (\ref{W17})) only contain Riemannian covariant
derivatives, and therefore, the 5-dimensional connection
\nabf{} is only an auxiliary quantity. Thus within some restrictions,
the 5-dimensional
connection coefficients $\Gamma^{\varepsilon }\,_{\mu\nu}$ can be
chosen arbitrarily. Therefore we choose the
5-dimensional connection on $\mathscr{M}_{5}$ in a certain way to make
calculations as simple as possible. First let us in general
assume:
\begin{eqnarray} \label{i.6}
        \nabf{\bm{\xi}}\bm{e}_{\varepsilon} =
        - \Upsilon_{\varepsilon}\,^{\nu}\bm{e}_{\nu} \qquad
        \qquad
        ( \Gamma^{\varepsilon  }\,_{\mu\nu}X^{\nu} =
        - \Upsilon^{\varepsilon}\,_{\mu}),
\end{eqnarray}
where $\Upsilon^{\varepsilon}\,_{\mu}$ is an arbitrary projector.
Taking into account the relation
\begin{eqnarray} \label{k.24}
\CR{\varepsilon }{\mu\nu} X^{\mu} = - g^{\varepsilon }\,_{\nu} +
\frac{1}{2}X_{\nu}\,^{\varepsilon },
\end{eqnarray}
which  follows immediately from(\ref{g.7}),  we obtain
\begin{eqnarray} \label{k.25}
          \sigma_{\mu\nu}\,^{\varepsilon }X^{\nu} =
          -\Sigma_{\mu}\,^{\varepsilon} -
           \frac{1}{2}X_{\mu}\,^{\varepsilon },
\end{eqnarray}
where we introduced the abbreviation
\begin{eqnarray} \label{k.15}
          \Sigma_{\mu}\,^{\varepsilon} \equiv\>  \Upsilon_{\mu}\,^{\varepsilon} -
          g_{\mu}\,^{\varepsilon}.
\end{eqnarray}
From the conditions (\ref{k.10}b) and (\ref{k.10}c)
we obtain the following relation for the torsion tensor:
\begin{eqnarray} \label{k.22}
          S_{\alpha\beta}\,^{\gamma} =
          A_{\alpha\beta}X^{\gamma}
          + \frac{1}{2S^{2}}h^{\gamma}\,_{\tau}
          (X_{\alpha}\Sigma_{\beta}\,^{\tau} -
           X_{\beta}\Sigma_{\alpha}\,^{\tau}).
\end{eqnarray}
Here we used the abbreviation
\begin{eqnarray} \label{k.19}
          A_{\varepsilon \tau} = S_{\varepsilon
          \tau}\,^{\mu}\zeta_{\mu}.
\end{eqnarray}
It is possible to show that the 5-dimensional connection has
the simplest form if according to (\ref{k.10}a) we put
\begin{eqnarray} \label{k.46}
       Q_{\varepsilon \alpha\beta}
        = \epsilon \sigma_{,\varepsilon } P_{\alpha\beta}.
\end{eqnarray}
In this case the quantity $\Sigma_{\mu\nu}$ will be antisymmetric
(with no other limitations):
\begin{eqnarray} \label{k.46a}
           \Sigma_{\mu\nu} = - \Sigma_{\nu\mu}.
\end{eqnarray}
The tensors $S_{\alpha\beta}\,^{\gamma}$ and
$\sigma_{\alpha\beta}\,^{\gamma}$ in this case are given by:
\begin{eqnarray} \label{k.47}
          S_{\alpha\beta}\,^{\gamma} =
          - \frac{X^{\gamma}}{2S^{2}}
          [X_{\alpha\beta} + (s_{\alpha}S_{,\beta} -
          s_{\beta}S_{,\alpha})]
          + \frac{1}{2S^{2}} (X_{\alpha}\Sigma_{\beta}\,^{\gamma}
          - X_{\beta}\Sigma_{\alpha}\,^{\gamma}),
\end{eqnarray}
\begin{eqnarray} \label{k.49}
   \sigma_{\lambda\mu\varepsilon } = &-& \frac{1}{2S}
   [(X_{\lambda\mu}s_{\varepsilon } - X_{\varepsilon \lambda}s_{\mu}
   + X_{\mu\varepsilon }s_{\lambda}) \nonumber \\ \bigskip\ds
 & + & 2s_{\mu}(s_{\lambda}S_{,\varepsilon } - s_{\varepsilon }S_{,\lambda})]
   - \frac{s_{\mu}}{S}\Sigma_{\lambda\varepsilon }\nonumber \\ \bigskip\ds
 & - & \frac{\epsilon }{2S}(P_{\lambda\mu}S_{,\varepsilon }
 - P_{\mu\varepsilon }S_{,\lambda } - P_{\lambda\varepsilon }S_{,\mu }).
\end{eqnarray}
From the last relation follows that the connection on
$\mathscr{M}_{5}$ has the simplest form if the quantity
$\Sigma_{\mu\nu}$ is defined according to (\ref{k.46a}) as follows:
\begin{eqnarray} \label{k.63}
\Sigma_{\varepsilon \nu} = G_{\varepsilon } X_{\nu} -  G_{\nu}X_{\varepsilon},
\end{eqnarray}
where the abbreviation (\ref{g.23}a) was used. Substituting the
last expression into the relations (\ref{k.47}) and (\ref{k.49}),
we obtain:
\begin{eqnarray} \label{k.64}
S_{\alpha\beta}\,^{\sigma} = s^{\sigma}\omega_{\alpha\beta},
\end{eqnarray}
\begin{eqnarray} \label{k.65}
\sigma_{\varepsilon \tau\nu} &=& \omega_{\varepsilon \tau}s_{\nu}
          - \omega_{\nu\varepsilon }s_{\tau}
          +\omega_{\tau\nu}s_{\varepsilon } \nonumber \\ \bigskip\ds
          &-&\frac{\epsilon }{2}(G_{\varepsilon }P_{\tau\nu}
          + G_{\tau}P_{\nu\varepsilon } -G_{\nu}P_{\tau\varepsilon }).
\end{eqnarray}
At the end of this section we would like to point out once again
that the connection on $\mathscr{M}_{5}$ is an intermediate
quantity. Its choice does not lead to any physical consequences. It
can be shown that for any choice of $\Sigma_{\varepsilon \alpha}$
and $Q_{\nu}\,^{\varepsilon \alpha}$ (these quantities have to
satisfy the conditions (\ref{k.10}) only) the 4-dimensional
physical equations get the same form. However, in the general case
all calculations become unwieldy. Therefore we don't present them
here fully; hereinafter we only will consider the case (\ref{k.46})
and (\ref{k.63}). Thus the torsion tensor
$S_{\alpha\beta}\,^{\varepsilon }$ and the tensor
$\sigma_{\alpha\beta}\,^{\varepsilon }$ take the simplest form,
i.e. (\ref{k.64}) and (\ref{k.65}), respectively.

\subsection{Projection of the Curvature Tensor and Related Quantities}

Now we have to analyse the equation (\ref{i.10}). In order to do
it, we can use the general relation
\begin{eqnarray} \label{k.84}
\til{T}_{\varepsilon \DD\mu\DD\lambda} -
\til{T}_{\varepsilon \DD\lambda\DD\mu} =
\til{T}_{\nu}G^{\nu}\,_{\varepsilon \mu\lambda} +
2\til{T}_{\varepsilon \DD\alpha}S_{\lambda\mu}\,^{\alpha},
\end{eqnarray}
where
\begin{eqnarray} \label{i.11}
G^{\alpha}\,_{\beta\gamma\delta}                =
         \Gamma^{\alpha}\,_{\beta\delta,\gamma} -
         \Gamma^{\alpha}\,_{\beta\gamma,\delta} +
\Gamma^{\alpha}\,_{\varrho\gamma}\Gamma^{\varrho}\,_{\beta\delta}-
\Gamma^{\alpha}\,_{\varrho\delta}\Gamma^{\varrho}\,_{\beta\gamma}.
\end{eqnarray}
To project the equation (\ref{k.84}) onto space-time we
need the following two relations:
\begin{eqnarray} \label{k.83}
\begin{array}{lll}\bigskip\ds
   \abc{a}
   \nabv{\beta}\nabv{\alpha}\til{T}_{\sigma}
   & = &
   \til{T}_{\delta\DD\varepsilon\DD \nu}P^{\nu}\,_{\beta}P^{\delta}\,_{\sigma}
   P^{\varepsilon }\,_{\alpha},\\ \bigskip\ds
   \abc{b}
   \nabv{\beta}\nabv{\alpha}\til{T}^{\sigma}
   & = &
   \til{T}^{\delta}\,_{\DD\varepsilon \DD\nu}P^{\nu}\,_{\beta}
   P^{\varepsilon }\,_{\alpha}h^{\sigma}\,_{\delta}.\\
\end{array}
\end{eqnarray}
The relation (\ref{k.83}a)  follows immediately from the equation
\begin{eqnarray} \label{c.74}
    \nabv{\beta}\nabv{\alpha}\til{T}_{\sigma}
    =
    (\til{T}_{\gamma\DD\tau}P^{\gamma}\,_{\delta}P^{\tau}\,_{\varepsilon})
    _{\DD\nu}P^{\nu}\,_{\beta}
    P^{\delta}\,_{\sigma}P^{\varepsilon }\,_{\alpha},
    \end{eqnarray}
in which the covariant derivatives
\begin{eqnarray} \label{k.72}
\abc{a}          s^{\mu}\,_{\DD\nu} = G^{\mu}s_{\nu},\quad
\abc{b}          s_{\mu\DD\nu} = G_{\mu}s_{\nu}
\end{eqnarray}
are substituted. The equation (\ref{k.83}b) can be similarly
proved. Proceeding further, we suppose that the vector
$\til{T}_{\varepsilon }$ satisfies the projector condition
(\ref{g.6}). The equation
\begin{eqnarray} \label{k.85}
     s^{\delta}\til{T}_{\varepsilon \DD\delta} = -(G^{\lambda}
     \til{T}_{\lambda})s_{\varepsilon }.
\end{eqnarray}
is fulfilled in this case. Using the relations (\ref{k.64}),
(\ref{k.72}) and (\ref{k.85}) we obtain the interesting equality
\begin{eqnarray} \label{k.86}
     \nabv{\beta}\nabv{\alpha}\til{T}_{\sigma}  -
     \nabv{\alpha}\nabv{\beta}\til{T}_{\sigma}
     =
     \til{G}^{\nu}\,_{\sigma\alpha\beta}\til{T}_{\nu},
\end{eqnarray}
where according to (\ref{th.11}) we used the abbreviation
\begin{eqnarray} \label{i.12}
    \til{G}^{\nu}\,_{\sigma\alpha\beta} \equiv\>
    h^{\nu}\,_{\mu}P^{\gamma}\,_{\sigma}P^{\delta}\,_{\alpha}
    P^{\varrho}\,_{\beta}G^{\mu}\,_{\gamma\delta\varrho}.
\end{eqnarray}
The equation (\ref{k.86}) being written in the basis
$\bm{e}_{n}$ (see (\ref{i.1}) ) is given by:
\begin{eqnarray} \label{k.86a}
T_{s;a;b} - T_{s;b;a} = T_{n}G^{n}\,_{sab},
\end{eqnarray}
where
\begin{eqnarray} \label{k.87}
   T_{n} = g^{\alpha}\,_{n}\til{T}_{\alpha},\quad
   G^{n}\,_{sab} = g^{n}\,_{\nu}g^{\sigma}\,_{s}g^{\alpha}\,_{a}
   g^{\beta}\,_{b}\til{G}^{\nu}\,_{\sigma\alpha\beta}.
\end{eqnarray}
As the equation (\ref{k.86a}) is correct for all space-time
vectors, the following relation is valid:
\begin{eqnarray} \label{k.89}
\riv^{n}\,_{sab} = G^{n}\,_{sab}.
\end{eqnarray}
Here $\riv^{n}\,_{sab}$ is the 4-dimensional Riemannian curvature
tensor
\begin{eqnarray} \label{i.13}
   \riv^{a}\,_{mnk}  \equiv\>
  \stackrel{4}{\CR{a}{mk}}_{,n} -
  \stackrel{4}{\CR{a}{mn }} _{,k } +
  \stackrel{4}{\CR{t}{mk }}
\stackrel{4}{\CR{a}{tn}}
    - \stackrel{4}{\CR{t}{mn }}
    \stackrel{4}{\CR{a}{tk}},
\end{eqnarray}
where we used the usual definition
\begin{eqnarray} \label{i.14}
   \stackrel{4}{\CR{k}{at}} \equiv\>
   \frac{1}{2}\ge^{ks}(\ge_{sa,t} +
   \ge_{ts,a} - \ge_{at,s}).
\end{eqnarray}
Obtaining the equation (\ref{k.86a}) we applied the
following relation:
\begin{eqnarray} \label{k.90}
   \nabv{\nu}\til{T}_{\mu}g^{\mu}\,_{m}g^{\nu}\,_{n} = T_{m;n}
   \quad
   \mbox{($T_{m} = g^{\nu}\,_{m}\til{T}_{\nu}$)}.
\end{eqnarray}

Let $T_{\mu}$ be an arbitrary one-form (covariant vector).
According to the projection formalism developed above we can
project this one-form onto the hypersurface $\mathscr{ M}_{4}$:
$T^{\mu} \longrightarrow \til{T}_{\mu}\equiv\>
P^{\alpha}\,_{\mu}T_{\alpha}$. Then the equations (\ref{k.23}) and
(\ref{k.83}) imply:
\begin{eqnarray} \label{k.95}
\nabv{\lambda}\nabv{\varepsilon}\til{T}_{\mu}
    =
   P^{\alpha}\,_{\mu}P^{\beta}\,_{\varepsilon}
   P^{\gamma}\,_{\lambda}
(\til{T}_{\alpha ;\beta} -   \sigma_{\alpha\beta}\,^{\nu}
\til{T}_{\nu})_{\DD\gamma}.
\end{eqnarray}
Substituting the relations (\ref{g.23}),
    (\ref{k.56}), (\ref{k.65}), (\ref{k.72}), (\ref{k.85}) as well as
    \begin{eqnarray} \label{k.93} a)\;
      P_{\mu\nu\DD\varepsilon } = \epsilon G_{\varepsilon }P_{\mu\nu}
      - s_{\varepsilon }(G_{\mu}s_{\nu}+ G_{\nu}s_{\mu}),\
b)\;
     P^{\nu}\,_{\mu\DD\varepsilon } = - s_{\varepsilon }(G^{\nu}s_{\mu}
     + G_{\mu}s^{\nu}),
\end{eqnarray}
\begin{eqnarray} \label{k.105}
     s^{\delta}\til{T}_{\varepsilon ; \delta} =
     s^{\delta}\til{T}_{\delta ; \varepsilon} =
 (\omega_{\varepsilon }\,^{\lambda} - s_{\varepsilon} G^{\lambda})
     \til{T}_{\lambda},
\end{eqnarray}
\begin{eqnarray} \label{k.107}
     G^{\varrho}\,_{\DD\varepsilon} = - \epsilon G_{\varepsilon }G^{\varrho}+
     g^{\varrho\lambda} G_{\lambda\DD\varepsilon }
\end{eqnarray}
into the last formula, we obtain the result
\begin{eqnarray} \label{k.110}
   \nabv{\lambda}\nabv{\varepsilon}\til{T}_{\mu}
  - \nabv{\varepsilon}\nabv{\lambda}\til{T}_{\mu}
& = &
    \til{T}_{\varrho}\Bigl\{\til{\rif}\;^
     {\varrho}\,_{\mu\varepsilon \lambda}
    - 2\omega_{\varepsilon \lambda}\omega_{\mu}\,^{\varrho} +
    \omega_{\mu\varepsilon }\omega_{\lambda}\,^{\varrho} -
    \omega_{\mu\lambda }\omega_{\varepsilon }\,^{\varrho}
\nonumber \\ \bigskip\ds
& - &
    \frac{\epsilon }{2}\Bigl[-\bigl(\nabv{\lambda}
         \til{G}_{\mu}\bigr)P^{\varrho}\,_{\varepsilon }
    + P_{\mu\varepsilon }g^{\varrho\alpha}\til{G}_{\alpha\DD\gamma}
      P^{\gamma}\,_{\lambda}
    + \bigl(\nabv{\varepsilon }\til{G}_{\mu}\bigr)P^{\varrho}\,_{\lambda }
\nonumber \\ \bigskip\ds
& - &
P_{\mu\lambda }g^{\varrho\alpha}\til{G}_{\alpha\DD\gamma}
      P^{\gamma}\,_{\varepsilon }\bigr)\Bigr]
    + \frac{\epsilon ^2}{4}\Bigl[-G^{\varrho}
     (G_{\varepsilon}P_{\mu\lambda}- G_{\lambda}
    P_{\mu\varepsilon }) \nonumber \\ \bigskip\ds
   & + & G_{\varepsilon }G_{\mu}P^{\varrho}\,_{\lambda}
 -  G_{\lambda }G_{\mu}P^{\varrho}\,_{\varepsilon }
\nonumber \\ \bigskip\ds
   & -& G^{\nu}G_{\nu}(P_{\mu\varepsilon }P^{\varrho}\,_{\lambda}
      - P_{\mu\lambda }P^{\varrho}\,_{\varepsilon })\Bigr]
\Bigr\}.
\end{eqnarray}
Now we are able to analyse the equation (\ref{i.10}a). However,
before doing it, let us summarize some formulas which are related
to the projection formalism.

In the 5-dimensional space $\mathscr{ M}_{5}$ the basis vectors and
basis one-forms were denoted by $\bm{e}_{\mu}$ and $\bm{e}^{\mu}$,
respectively. The 4-dimensional holonomic hypersurface $\mathscr{
M}_{4}$ in the 5-dimensional space $\mathscr{M}_{5}$ is identified
with the 4-dimensional space-time. The quantities projected onto
the hypersurface $\mathscr{ M}_{4}$ were denoted by a tilde (see
(\ref{th.11})). The quantities $x^{i}$ parametrizing curves of the
congruence (\ref{th.8}) can be used as coordinates in the
space-time. The tangent vectors $\bm{e}_{i}$ to the coordinate
lines ($\bm{e}_{i} = \pderiv{}{x^{i}}$) of this 4-dimensional
coordinate system form a 4-dimensional vector space $\til{T}_{P}$
(see (\ref{g.58}). Between the basis vectors $\bm{e}_{i}$ and
$\bm{e}_{\mu}$ exists consistency (\ref{g.56}). Similar relations
are valid for the dual basis $\bm{e}^{i}$ ($\bm{e}^{i} = dx^{i}$),
too. Thus the 4-dimensional vectors and one-forms can be rewritten
in the following form:
\begin{eqnarray} \label{i.15}
\begin{array}{ll}\bigskip\ds
a)\ &
   \til{\bm{V}} = \til{V}^{\alpha}\bm{e}_{\alpha} =
   V^{i}\bm{e}_{i}\qquad  (\til{V}^{\alpha}
   = g^{\alpha}\,_{i}V^{i}, \
   V^{i} = g^{i}\,_{\alpha}\til{V}^{\alpha})
   \\ \bigskip\ds
b)\ &
   \til{\bm{\omega}} = \til{\omega}_{\alpha}\bm{e}^{\alpha} =
   \omega_{i}\bm{e}^{i}\qquad  (\til{\omega}_{\alpha}
   = g^{i}\omega^{i}\,_{\alpha}, \
   \omega_{i} = g^{\alpha}\,_{i}\til{\omega}_{\alpha}).\\
\end{array}
\end{eqnarray}
Let us remember that on the hypersurface $\mathscr{ M}_{4}$ we
introduced two metrics: the induced metric $\til{\bm{g}}$ and the
physical metric $\stackrel{4}{\bm{g}}$. These metrics are connected
by means of the relation (\ref{k.1b}). As the physical metric
differs in general from the induced one, one should be careful in
defining 4-dimensional physical quantities.

Using the abbreviation
\begin{eqnarray} \label{k.113a}
     \om_{mn} \equiv\>  \omega_{mn} =
     g^{\mu}\,_{m}g^{\nu}\,_{n}\til{\omega}_{\mu\nu}=
     g^{\mu}\,_{m}g^{\nu}\,_{n}\omega_{\mu\nu}
\end{eqnarray}
we obtain from (\ref{i.15})  and  (\ref{k.1b}) the following
relations
\begin{eqnarray} \label{k.113b}
\abc{a}
    \omega_{m}\,^{n} \equiv\>  g^{\mu}\,_{n}g^{n}\,_{\nu}
    \til{\omega}_{\mu}\,^{\nu} = e^{\epsilon \sigma}
    \om_{m}\,^{n},\quad
\abc{b}
    \omega^{mn} \equiv\>  g^{m}\,_{\mu}g^{n}\,_{\nu}
    \til{\omega}^{\mu\nu} = e^{2\epsilon \sigma}
    \om^{mn},
\end{eqnarray}
    where
     $\om_{m}\,^{n} = \ge^{nk}
    \om_{mk}$ and $\om^{mn} =
    \ge^{mk}\ge^{nl}
    \om_{kl}$.
In a similar way we deduce from (\ref{g.64}) and (\ref{g.23}a) the
equations
\begin{eqnarray} \label{k.112}
a)\
   g^{\mu}\,_{m}G_{\mu} = - g^{\mu}\,_{m}\sigma_{,\mu} = -
   \sigma_{,m} \>, \quad
b)\
   g^{m}\,_{\mu} \til{G}^{\mu} = - \til{g}^{\mu\nu}g^{m}\,_{\nu}
   \sigma_{,\mu} = - e^{\epsilon \sigma}\sigma^{,m},
\end{eqnarray}
where  $\sigma^{,m}  \equiv\>   \ge^{mn}\sigma_{,n}$.

Let us note that the space-time indices are to be moved with the
help of the space-time metric $\stackrel{4}{\bm{g}}$.

It can be shown that for the arbitrarily projected quantities
$\til{\bm{\omega}}$ and $\til{\bm{V}}$ the relation
\begin{eqnarray} \label{i.16}
     \til{\omega}_{\tau}\til{V}^{\tau} = \omega_{n}V^{n}\quad
     \mbox{($\omega_{m}\equiv\>  g^{\mu}\,_{m}\til{\omega}_{\mu}, \
     V^{m} \equiv\>  g^{m}\,_{\tau}\til{V}^{\tau}$)}
\end{eqnarray}
is true. From (\ref{k.86}) and (\ref{k.89}) follows
\begin{eqnarray} \label{k.114}
     g^{\sigma}\,_{s}g^{\alpha}\,_{a}g^{\beta}\,_{b}
     \bigl(
     \nabv{\beta}\nabv{\alpha}\til{T}_{\sigma} -
     \nabv{\alpha}\nabv{\beta}\til{T}_{\sigma} \bigr)
     = \riv^{n}\,_{sab}\stackrel{4}{T}_{n}\quad
     \mbox{($\stackrel{4}{T}_{n}\equiv\>  g^{\sigma}\,_{n}
     \til{T}_{\sigma}$).}
\end{eqnarray}

Using the relations (\ref{k.110}) and further the relations
(\ref{k.112}) to (\ref{k.114}), we obtain the final result
\begin{eqnarray} \label{k.121}
    \til{\rif}\;^{a}\,_{mkl} & \equiv\>  &
    g^{a}\,_{\alpha}g^{\mu}\,_{m}g^{\varepsilon }\,_{k}g^{\lambda}\,_{l}
    R^{\alpha}\,_{\mu\varepsilon \lambda} = \nonumber\bigskip\\ \ds
& = &
    \riv^{a}\,_{mkl} +
    e^{\epsilon \sigma}\left(2\om_{kl}\om_{m}\,^{a}-\om_{mk}\om_{l}\,^{a}
    + \om_{ml}\om_{k}\,^{a}\right)\nonumber\bigskip\\ \ds
& + & \frac{\epsilon }{2}\left(\ge^{a}\,^{k}\sigma_{,m;l}
    - \ge_{mk}\sigma^{,a}\,_{;l} -\ge^{a}\,_{l}\sigma_{,m;k}
    +\ge_{ml}\sigma^{,a}\,_{;k}\right)\nonumber\bigskip\\ \ds
& - & \frac{\epsilon ^2}{4}
    \Bigl[-\sigma^{,a}\left(\sigma_{,k}\ge_{ml}-\sigma_{,l}\ge_{mk}\right)
    + \ge^{a}\,_{l}\sigma_{,k}\sigma_{,m}\nonumber\bigskip\\ \ds
& - &
    \ge^{a}\,_{k}\sigma_{,m}\sigma_{,l} -
    (\sigma_{,c}\sigma^{,c})\left(\ge_{mk}\ge^{a}\,_{l}
    - \ge_{ml}\ge^{a}\,_{k}\right)\Bigr].
\end{eqnarray}
In order to find the projection of the 5-dimensional Ricci tensor
onto space-time $\mathscr{ M}_{4}$ let us consider the relation
\begin{eqnarray} \label{k.122}
    \til{\rif}_{mn} \equiv\>  g^{\mu}\,_{m}g^{\nu}\,_{n}
    \til{\rif}_{\mu\nu} = g^{\mu}\,_{m}g^{\nu}\,_{n}\Bigl[
    \til{\rif}{}^{\varrho}\,_{\mu\nu\varrho} +
    P^{\alpha}\,_{\mu}P^{\beta}\,_{\nu} (s_{\varrho}s^{\sigma}
    \rif^{\varrho}\,_{\alpha\beta\sigma})\Bigr],
 \end{eqnarray}
where
\begin{eqnarray} \label{k.122a}
    \til{\rif}_{\mu\nu} \equiv\>  P^{\alpha}\,_{\mu}P^{\sigma}\,_{\nu}
    \til{\rif}_{\alpha\sigma},\quad
    \til{\rif}{}^{\varrho}\,_{\mu\nu\sigma} \equiv\>
    h^{\varrho}\,_{\lambda}P^{\alpha}\,_{\mu}P^{\beta}\,_{\nu}
    P^{\tau}\,_{\sigma} \rif^{\lambda}\,_{\mu\nu\sigma\tau}.
\end{eqnarray}
From (\ref{k.86}), (\ref{k.89}) and (\ref{k.110}) we find
\begin{eqnarray} \label{k.124}
    g^{\mu}\,_{m}g^{\nu}\,_{k}
    \til{\rif}{}^{\varrho}\,_{\mu\nu\varrho} & = &
    \riv_{mk} + 3 e^{\epsilon \sigma}\om_{ka}\om_{m}\,^{a}
    -\frac{\epsilon }{2}
         \left(\sigma_{,m;k} + \frac{1}{2}\ge_{mk}\sigma^{,a}\,_{;a}\right)
    \nonumber\bigskip\\ \ds
   & + &
    \frac{\epsilon ^2}{2}\left[\ge_{mk}(\sigma^{,a}\sigma_{,a}) -
    \sigma_{,k}\sigma_{,m}\right].
\end{eqnarray}
The second term on the right hand side of the equation
(\ref{k.122}) can be calculated in the simplest way using the
formulas (\ref{k.24}) and (\ref{k.59}). The result is
\begin{eqnarray} \label{k.125}
     \rif^{\sigma}\,_{\alpha\beta\varrho}s_{\sigma}s^{\varrho}
     = \sigma_{,\beta;\alpha} + \sigma_{,\alpha}\sigma_{,\beta}
     -\omega_{\lambda\beta}\omega^{\lambda}\,_{\alpha}+
     \omega_{\lambda[\beta}s_{\alpha]}\sigma^{,\lambda}-
     s_{\alpha}s_{\beta}\sigma^{,\lambda}\sigma_{,\lambda}.
\end{eqnarray}
By substituting the expressions (\ref{k.124}) and (\ref{k.125})
into (\ref{k.122}) we get the formula
\begin{eqnarray} \label{k.127}
     \rif_{mn} &=& \riv_{mn} + 2e^{\epsilon \sigma}\om_{na}\om_{m}\,^{a}
     - \frac{\epsilon }{2}\ge_{mn}\sigma^{,a}\,_{;a}
    +(1+\epsilon -\frac{\epsilon ^2}{2})\sigma_{,m}\sigma_{,n}
         \nonumber \\ \ds &+&
         (1-\epsilon )\left[
         \sigma _{,m;n}-\frac{\epsilon }{2}\ge_{mn}\sigma^{,a}\,_{;a}
         \right].
\end{eqnarray}
Here we used the equation
\begin{eqnarray} \label{k.126}
    P^{\alpha}\,_{\mu}P^{\beta}\,_{\nu}\sigma_{,\beta;\alpha}
    = - \nabv{\mu}\til{G}_{\nu} +
    \frac{\epsilon }{2}[2\til{G}_{\nu}\til{G}_{\mu} -
    (G^{\varrho}\til{G}_{\varrho})P_{\mu\nu}].
\end{eqnarray}
By means of the expression
\begin{eqnarray} \label{k.128}
     \rif = (P^{\mu\nu} + s^{\mu}s^{\nu})\rif_{\mu\nu}
\end{eqnarray}
and taking into account the intermediate formulas
\begin{eqnarray} \label{k.129}
           P^{\mu\nu} \rif_{\mu\nu} =
           e^{\epsilon \sigma}\ge^{mn}\til{\rif}_{mn},
\end{eqnarray}
\begin{eqnarray} \label{k.129a}
             s^{\mu}s^{\nu}\rif_{\mu\nu} =
             \sigma^{,\alpha}\,_{;\alpha}-
             \omega_{\lambda\alpha}\omega^{\lambda\alpha},
\end{eqnarray}
\begin{eqnarray} \label{k.130}
     \sigma^{,\alpha}\,_{;\alpha} = e^{\epsilon \sigma}
          \left[
                \sigma^{,i}\,_{;i}+(1-\epsilon)\sigma ^{,i}\sigma _{,i}
          \right],
\end{eqnarray}
we find the  result
\begin{eqnarray} \label{k.131}
     \rif = e^{\epsilon \sigma}
          \Bigl(\riv + e^{\epsilon \sigma}\om_{mn}\om^{mn}
     - (2-3\epsilon )\sigma^{,m}\,_{;m}+
          2(1-\epsilon +\frac{3}{4}\epsilon ^2)\sigma^{,m}\sigma_{,m}
     \Bigr).
\end{eqnarray}

Now we immediately can obtain the 4-dimensional field equations of
PUFT being restricted to the versions II and III of PUFT. Today the
version I of PUFT has only historical value.

\subsection{Version II}

By projecting the 5-dimensional field equations (\ref{W1}) onto the
4-dimensional space-time with the help of the projection formalism
developed above we obtain the 4- dimensional field equations of
PUFT. As the space-time metric $\bm{\ge}$ is connected with the
induced metric $\til{\bm{g}}$ on the hypersurface $\mathscr{M}_4$
by means of (\ref{k.1}), in case of the version II of PUFT it is
necessary to put:
\begin{eqnarray}\label{z4}
\left.
\begin{array}{lcl} \ds
G_{\mu  \nu }&=&
                    \stackrel{5}{R}_{\mu\nu}-
    \frac{1}{2}g_{\mu\nu}\stackrel{5}{R}{}+
         \lambda_0 S_0 e^{\sigma }
    \left(g_{\mu \nu }+s_{\mu }s_{\nu }\right)\\  \ds
\epsilon &=&1 \\
\end{array} \right\}.           
\end{eqnarray}

\subsubsection{Generalized  gravitational field equation}

Using the last results obtained from the equation (\ref{i.10}a)
within the framework of(\ref{z4}), the field equations read:

\begin{eqnarray} \label{k.138}
     \riv_{mn} - \frac{1}{2}\ge_{mn}\riv
          + \lambda_0 S_0 \ge_{mn} = \kappa_{0}
     \stackrel{4}{T}_{mn},
\end{eqnarray}
where
\begin{eqnarray} \label{k.139}
   \stackrel{4}{T}_{mn} = \stackrel{4}{\theta}_{mn}
   + \frac{1}{\kappa_{0}}\Bigl[2e^{\sigma}(\om_{ma}
     \om^{a}\,_{n}
   + \frac{1}{4}\ge_{mn}\om_{ab}\om^{ab}) \nonumber\bigskip\\ \ds
   - \frac{3}{2}(\sigma_{,m}\sigma_{,n} - \frac{1}{2}
   \ge_{mn}\sigma^{,a}\sigma_{,a})\Bigr]
\end{eqnarray}
and
\begin{eqnarray} \label{k.137}
 \stackrel{4}{\theta}_{mn} \equiv\>   g^{\mu}\,_{m}g^{\nu}\,_{n}
\til{\stackrel{5}{\theta}}_{\mu\nu}.
\end{eqnarray}

\subsubsection{Generalized electromagnetic field equations
(Maxwell equations)}

Comparing the relation (\ref{k.139}) with the expression
(\ref{th.3}), we find that the angular velocity of the congruence
(\ref{th.8}) $\omega_{\alpha\sigma}$ is connected with the
electromagnetic strength tensor in the following way:
\begin{eqnarray} \label{k.145}
       B_{\alpha\sigma} \equiv\>  \til{B}_{\alpha\sigma}=
       B_{0}e^{a\sigma}\omega_{\alpha\sigma},
\end{eqnarray}
where the constant $B_{0}$ depends on the system of units. We choose the
constant $a$ in order to fulfill the next equation
\begin{eqnarray} \label{i.17}
      B_{<\mu\nu\DD\alpha>} = 0 .
\end{eqnarray}
It is easy to see that the relation
\begin{eqnarray} \label{k.146}
     B_{<\mu\nu\DD\alpha>}  = B_{<\mu\nu;\alpha>}
\end{eqnarray}
holds. Using the expression (\ref{k.59}) and the equation
\begin{eqnarray} \label{i.18}
     X_{<\mu\nu;\alpha>} = 0
\end{eqnarray}
we find
\begin{eqnarray} \label{i.19}
     B_{<\mu\nu;\alpha>} =
     B_{0}e^{a\sigma}(1+a)\;\omega_{<\alpha\mu}\sigma_{,\nu>}.
\end{eqnarray}
This implies  $a=-1$ and
\begin{eqnarray} \label{k.147}
       B_{\alpha\sigma} =
       B_{0}e^{-\sigma}\omega_{\alpha\sigma},
       \qquad       B_{mn} \equiv\>  g^{\mu}\,_{m} g^{\nu}\,_{n}
       B_{\mu\nu}=
       B_{0}e^{-\sigma}\om_{mn}.
\end{eqnarray}
It is obvious (see (\ref{k.2}) and (\ref{i.1})) that the
electromagnetic field strength tensor satisfies the cyclic Maxwell
system
\begin{eqnarray} \label{i.19a}
          B_{<mn;k>} = 0.
\end{eqnarray}
By substituting (\ref{k.147}) in the expression (\ref{k.139}) we
find that the electromagnetic induction tensor is to be defined as
follows:
\begin{eqnarray} \label{i.20}
                       H_{mn} = e^{3\sigma}B_{mn},
\end{eqnarray}
and the constant $B_{0}$ can be chosen as
\begin{eqnarray} \label{i.21}
                  B_{0} = \pm\sqrt{\frac{8\pi}{\kappa_{0}}}.
\end{eqnarray}
In this case the electromagnetic part of the energy-momentum tensor
$\stackrel{4}{T}_{mn}$ (\ref{k.139}) takes its usual form (in the
Gaussian system of units):
\begin{eqnarray} \label{i.22}
           E_{mn}={1\over 4\pi}(B_{mk}H^{k}\,_{n}+{1\over 4}
           \ge_{mn}B_{jk}H^{jk}).
\end{eqnarray}

It is easy to see that the one-form
\begin{eqnarray} \label{i.23}
          A_{\mu} \equiv\>  B_{0}S_{0}
          P^{\sigma}\,_{\mu}\zeta_{\sigma}
          = B_{0}S_{0}P^{\sigma}\,_{\mu}\tau_{,\sigma}
\end{eqnarray}
has the following properties:
\begin{eqnarray} \label{i.24}
       A_{\mu\DD\nu} - A_{\nu\DD\mu}
       = B_{\nu\mu} \quad \mbox{and}\quad
       A_{m,n} - A_{n,m}
       = B_{nm} \quad
       \mbox{($A_{m}= g^{\sigma}\,_{m}A_{\sigma}$)}.
\end{eqnarray}
Thus the orthogonal vector, projected into the hypersurface
$\mathscr{ M}_{4}$ in an appropriate way, is the electromagnetic
vector potential.

Now we are ready to expound the equation (\ref{i.10}b). The result
is
\begin{eqnarray} \label{k.154}
     H^{mn}\,_{;n} = \frac{4\pi}{c}j^{m},
\end{eqnarray}
where the abbreviations we used are given by
\begin{eqnarray} \label{i.25}
       j^{m} = \frac{\kappa_{0}B_{0}c}{4\pi }
       \stackrel{4}{\theta}{}^{m}
       \quad \mbox{and}\quad
       \stackrel{4}{\theta}{}^{m} \equiv\>  \ge_{mn}g^{\mu}\,_{n}
       \til{\stackrel{5}{\theta}}_{\mu}.
\end{eqnarray}

\subsubsection{Field equation of the scalaric field $\sigma$}

Using the Relations (\ref{k.129a}) and (\ref{i.9}) we can rewrite
the equation (\ref{i.10}c) in the form
\begin{eqnarray} \label{k.150}
     \sigma^{,m}\,_{;m} = \frac{\kappa_{0}}{8\pi}B_{mn}H^{mn}
     + \frac{2}{3}\kappa_{0}\vartheta,
\end{eqnarray}
where the following definition was used:
\begin{eqnarray} \label{i.26}
          \vartheta \equiv\>  \stackrel{5}{\theta}e^{-\sigma}
          - \frac{1}{2} \ge^{mn}\stackrel{4}{\theta}_{mn}.
\end{eqnarray}

\subsection{Version III}

In the version III of PUFT the space-time metric $\bm{\ge}$
coincides with the on the hypersurface $\mathscr{M}_4$ induced
metric $\til{\bm{g}}$ (see (\ref{k.1a})). This makes the projection
formalism a little bit easier, because $\epsilon =0$. On the
contrary the 5-dimensional field equations (\ref{W17}) are more
complicated. Further investigations have shown that the case
$\kappa_0 K_0=-2$, already mentioned above, is of a particular
interest. Hence for the version III we find the following equation
\begin{eqnarray}\label{z5}
\left.
\begin{array}{lcl} \ds
G_{\mu\nu}&=&
   R_{\mu\nu}-{1\over 2} g_{\mu\nu} \stackrel{5}{R} -
  {1\over S} S_{,\mu ;\nu} -
  {2 \over {S^2}} S_{,\mu}S_{,\nu}\\&-& {1\over S}
  s_{\mu}s_{\nu} \left(4S^{,\tau}_{\ ;\tau}
 -{6 \over S} S_{,\tau} S^{,\tau} + {3 \lambda_0 \over S}- {S \over
  2} \stackrel{5}{R} \right)
  + {1 \over S} g_{\mu \nu}\left(
  S^{,\tau}_{\ ;\tau}
+ {\lambda_c \over S}\right)\\
\epsilon &=& 0
\end{array}
\right\}.
\end{eqnarray}
Following the above introduced procedure of deducing the field
equations, one obtains the system of equations listed below
\cite{Schmutzer6}.

\subsubsection{Generalized gravitational field equation}

                \begin{eqnarray} \label{k.138a}
     \riv_{mn} - \frac{1}{2}\ge_{mn}\riv
          + \frac{\lambda_0}{ S_0^2}e^{-2\sigma} \ge_{mn} = \kappa_{0}
     \stackrel{4}{T}_{mn},
\end{eqnarray}
where
$\ds \stackrel{4}{T}_{mn}=\theta _{mn} +E_{mn}+S_{mn}$
 with
 \begin{eqnarray} \label{i.22a}
           E_{mn}={1\over 4\pi}(B_{mk}H^{k}\,_{n}+{1\over 4}
           \ge_{mn}B_{jk}H^{jk}),\quad
    S_{mn} =  \frac{2}{\kappa_0}
    (\sigma_{,m}\sigma_{,n}-{1\over 2}g_{mn}
    \sigma_{,k}\sigma^{,k})
\end{eqnarray}
holds.
$\ds
\stackrel{4}{\theta}_{mn} \equiv\>   g^{\mu}\,_{m}g^{\nu}\,_{n}
\til{\stackrel{5}{\theta}}_{\mu\nu}$
is the energy-momentum tensor of the substrate.

\subsubsection{Generalized electromagnetic field equations
(Maxwell equations)}

 \begin{eqnarray} \label{h.5a}
           a)\; H^{mn}\,_{;n}={4\pi\over c}j^{m} , \qquad
          b)\; B_{[mn,k]}=0, \qquad
          c)\; H_{mn}=e^{2\sigma}B_{mn},
\end{eqnarray}
where we used the abbreviations
\begin{eqnarray}\label{h.5b}
\abc{a} B_{mn} \equiv\>  g^{\mu}\,_{m} g^{\nu}\,_{n}
       B_{\mu\nu}= B_0
       e^{-\sigma}\om_{mn},\quad
\abc{b}
j^{m} = \frac{\kappa_{0}B_{0}c}{4\pi }
       \stackrel{4}{\theta}{}^{m} e^{\sigma}
 \end{eqnarray}
and $\ds
        \stackrel{4}{\theta}{}^{m} \equiv\>  \ge_{mn}g^{\mu}\,_{n}
       \til{\stackrel{5}{\theta}}_{\mu}
                 $, $
                B_{0} = \pm\sqrt{\frac{8\pi}{\kappa_{0}}}$.

\subsubsection{Field equation of the scalaric field $\sigma$}

 \begin{eqnarray} \label{k.150a}
     \sigma^{,m}\,_{;m} = -\frac{\kappa_{0}}{16\pi}B_{mn}H^{mn}
     - \frac{\kappa_{0}}{2}\vartheta-
          \frac{\lambda_0}{S_0^2}e^{-2\sigma}
          \quad \mbox{with} \quad
          \vartheta \equiv\>  \stackrel{5}{\theta}
          - \ge^{mn}\stackrel{4}{\theta}_{mn}.
\end{eqnarray}
\section{Concluding Notes }
Now let us summarize the basic ideas of the new geometrical
approach to the axiomatics of Schmutzer's 5-dimensional Projective
Unified Field Theory.

The mathematical basis for the 5-dimensional Projective Unified
Field Theory forms the group of all 5-dimensional homogeneous
coordinate transformations of degree one (\ref{2.1}). The
5-dimensional geometry, constructed on this group, supposes the
existence of a Killing vector field. The integral curves of this
vector field form a Killing congruence (\ref{th.8}) which is the
basis of the projection formalism developed here. The angular
velocity $\omega_{\mu\nu}$ of this congruence is interpreted as the
electromagnetic field strength tensor (see (\ref{k.147}) and
(\ref{h.5a})). It is well known that, if $\omega_{\mu\nu}=0$ holds,
a hypersurface, holonomic and orthogonal to the congruence exists.
There are two possibilities to construct an axiomatics of PUFT:
abandoning either
\textbf{holonomicity} or
\textbf{orthogonality}. The first of the two possibilities
was investigated in detail in numerous papers by Schmutzer (see
\cite{Schmutzer1,Schmutzer5,Schmutzer6} and there quoted papers).
The second possibility was considered in the present paper. In this
case it is possible to say that the \textbf{holonomicity} of
space-time and the \textbf{non-orthogonality} of the given
congruence with respect to the space-time hypersurface are embodied
in the basis of the axiomatics offered here. In this way PUFT has
got a new geometrical interpretation.

Inner curvature of space-time ($\mathscr{M}_4$) identified with the
hypersurface $\tau = \mbox{const}$ (see page \pageref{Hyper})
describes the gravitation. The norm of the Killing vector field
$\bm{\xi}$ (\ref{th.9}) is connected with the new scalaric field
$\sigma$: $\sqrt{\xi^{\nu} \xi_{\nu}}=S_0e^{\sigma}$. The tensor of
the angular velocity $\omega_{\mu\nu}$ (\ref{g.15}) of the Killing
congruence (\ref{th.8}) describes the electromagnetic field. Thus
the orthogonal vector projected in an appropriate way onto the
hypersurface $\tau=\mbox{const}$ ($\mathscr{M}_4$) is the
electromagnetic vector potential (see (\ref{i.23}) and
(\ref{i.24})). The relation (\ref{i.23}) implies that the
electromagnetic potential vanishes if the hypersurface
$\mathscr{M}_4$ is orthogonal to congruence (\ref{th.8}).

It is easy to show that physically the 4-dimensional field
equations in the version II of PUFT differ slightly from the
corresponding equations in the version III \textbf{physically}
slightly. Here we won't dwell on this problem, therefore let us
just remark that the cosmological term in the equation (\ref{W1})
can be accepted in the form $\Lambda _{\mu \nu } = \lambda _0
e^{-\sigma} P_{\mu \nu }$ ($\Lambda ^{\mu}\,_{ \nu ;\mu }=0$) as
well. In this case additional terms containing a cosmological
constant in the equations (\ref{k.138}) and (\ref{k.150}) take the
form $\lambda _0 e^{-2\sigma }\ge_{mn}$ and $\frac{4}{3}\lambda _0
e^{-2\sigma }$, respectively.

In conclusion let us emphasize that the axiomatics constructed here
leads to the same 4-dimensional field equations which formerly were
obtained by E.Schmutzer in a different way.

\hrulefill
\section{Appendix. Results of Application of PUFT}

Since 1995 a series of papers by E.Schmutzer on a closed
homogeneous isotropic cosmological model of the universe and on the
influence of the expansion of such a model on cosmogony and
astrophysics appeared ( see \cite{Schmutzer6} where further
literature is quoted) or are in press \cite{Schmutzer7}. Let us
mention some main results:
\begin{itemize}
\item
  In order to be in agreement with the equivalence principle
the usual concept of mass is basically changed: mass depends on the
cosmological scalaric field. Hence follows a considerable change of
the cosmological situation at the start of the universe (fulfilling
of certain aspects of Mach's principle).
\item
The big bang singularity does not exist. The "big start" (Urstart)
of the universe begins softly and is (using a certain physically
motivated choice of parameters) characterized by a kind of
oscillations: expansion interrupted by small contractions.
\item
The cosmological scenario appears to be divided into a short
repulsion (antigravitational) era (duration of 128 years) and a
cosmologically long attraction era (age of the universe = 18
billions of years).
\item
The Hubble factor ("constant") is $75 \mbox{km}/s\;\mbox{Mpc}$.
\item
Maxima and minima in the curves for the temporal behaviour of the
cosmological mass density and the temperature could be interesting
for the explanation of cosmogonic activities (birth of galaxies and
stars).
\item
The equation of motion of a body is in full agreement with the
Einstein effects (periastron motion, deflection and frequency shift
of electromagnetic waves).
\item
Further consequences of the equation of motion are:
\begin{itemize}
\item
Time dependence of the "effective gravitational constant" with the
present relative value: $3.5\cdot 10^{-11}/\mbox{year}$.
\item
For an orbiting body around a center: positive value of the angular
(secular) acceleration, negative values of the time derivatives of
the orbital radius (decrease), revolution period (decrease),
excentricity (transition from elliptic to cyclic orbits).
\item
Heat production in a moving body with application to the moon,
planets, sun, galaxy etc. with remarkably interesting numerical
results.
\end{itemize}
\end{itemize}

\vspace{2ex}
\hrulefill

\vspace{2ex}\sf
The author would like to thank Professor Ernst Schmutzer for numerous
helpful discussions on the axiomatics of PUFT.


\begin{thebibliography}{99}
\bibitem{Kaluza}
    Th.Kaluza. {\it  Sitzungsber.  d.  preuss.  Akad.  d.  Wiss.,
    Phys.-math. Kl.} 541 (1921)
\bibitem{Klein}
O.~Klein. {\it Z. Phys.} {\bf 37}, 895 (1926)
\bibitem{Dantzig}
         D.~van Dantzig. {\it Math. Ann.} {\bf 106}, 400 (1932)
\bibitem{Veblen}
         O.~Veblen. {\it Projektive  Relativitaetstheorie},  {\sl
    Springer-Verlag}, Berlin 1933
\bibitem{Jordan}
    P.~Jordan.  {\it  Schwerkraft  und  Weltall},  {\sl    Vieweg
    Braunschweig} 1955
\bibitem{Schmutzer1}
         E.~Schmutzer. {\it  Relativistische    Physik},    {\sl    Teubner
         Verlagsgesellschaft}. Leipzig 1968
\bibitem{Schouten}
         J.A.~Schouten. {\it  Ricci Calculus},   {\sl    Springer
    Verlag}. Berlin-Goettingen-Heidelberg 1954
\bibitem{Schmutzer2}
         E.~Schmutzer. {\it Proceedings of the 9th International Conference
         on General Relativity and Gravitation}, {\sl Deutscher Verlag  der
         Wissenschaften}, Berlin 1983
\bibitem{Wesson1}
P.S.~Wesson. \textit{Space -- Time -- Matter. Modern
Kaluza--Klein Theory.} {\sl World Scientific.}
Singapore-New Jersey-London-Hong Kong 1999
\bibitem{Vladimirov}
        Ju.S.~Vladimirov
                  \textit{Relational theory of space-time and interactions.
                  Volume 2. Theory of physical interactions.}
                  \textsl{Moscow State Univ. Press.} Moscow 1998
\bibitem{Wesson}
        J.M~Overduin and P.S.~Wesson.  {\it  Physics  Reports}  {\bf
    283}, 303 (1997)
\bibitem{Wesson2}
P.S.~Wesson, B.~Mashhoon and H.~Liu.
\textit{Phys. Letters} \textbf{B 456},34-37 (1999)
 \bibitem{Schmutzer5}
         E.~Schmutzer.
 {\it Fortschr.
         Phys.} {\bf 43}, 613 (1995)
\bibitem{Hawking}
         S.W.~Hawking and G.F.R.~Ellis.  {\it  The  Large  Scale  Structure  of
         Space--Time.} {\sl Cambridge University Press.}
\bibitem{Schmutzer6}
         E.~Schmutzer. {\it Astronomische Nachrichten.}
                        \textbf{320},1 (1999)
\bibitem{Schmutzer7}
E.~Schmutzer. {\it Jahrbuch 1999 der Deutschen Akademie der
Naturforscher Leopoldina (Halle/Salle)} \textbf{45}, 65 (2000);
{\it Astronomische Nachrichten} (submitted);
{\it General Relativity and Gravitation} [special issue] (submitted)
\end{thebibliography}
    \end{document}